\documentclass[12pt]{article}
\usepackage{amsmath,amssymb,latexsym}
\usepackage{theorem}

\usepackage{authblk}

\newtheorem{definition}{Definition}
\newtheorem{lemma}{Lemma}

\newtheorem{corollary}{Corollary}
\newtheorem{theorem}{Theorem}
\newtheorem{proposition}{Proposition}

\newtheorem{open-probl}{Open problem}

\newtheorem{exampleA}{Example}
\newtheorem{remarkA}{Remark}
\newenvironment{example}
{\begin{exampleA}\begin{normalfont}}{\end{normalfont}\end{exampleA}}
\newenvironment{remark}
    {\begin{remarkA}\begin{normalfont}}{\end{normalfont}\end{remarkA}}
\newenvironment{proof}[1][Proof.]
    {\addvspace{\bigskipamount}\noindent\emph{#1} }
    {\par\addvspace{\bigskipamount}}

 \newcommand{\G}{C}
 \newcommand{\closure}{\mathbf Cl}
 \newcommand{\Closure}[1]{{\mathbf Cl}(#1)}

\newcommand{\N}{\mathbb{N}}

\newcommand{\Q}{\mathbb{Q}}
\newcommand{\Z}{\mathbb{Z}}
\newcommand{\R}{\mathbb{R}}

\newcommand{\card}{\mathop{\mathrm{Card}}\nolimits}

\newcommand{\FIN}{{\cal L}(IND_{FIN})}
\newcommand{\META}{{\cal L}(Meta_{Lin})}

\newcommand{\derivestep}[1][]{\mathrel{\mathop{\Rightarrow}_{#1}}}

\newcommand{\qed}{\ifmmode\raisebox{3.3pt}{\fbox{}}\else{\ifhmode
\unskip\fi\nobreak\hfil\penalty50\hskip1em\null\nobreak\hfil\raisebox
{3.3pt}{\fbox{}}\parfillskip=0pt\finalhyphendemerits=0\endgraf}\fi}

\title{On the Intersection Problem for Quantum  Finite Automata\thanks{ The second and the third author have been partially supported by Sapienza Ateneo 2023 Project ``Representation Theory and Applications''.
    The authors acknowledge their membership to the National Group for Algebraic and Geometric Structures, and their Applications (GNSAGA--INdAM)}
 }

\newlength{\transwidth}
\newcommand{\trans}[1]{\settowidth{\transwidth}{$\;_{#1}\;$}
\stackrel{#1}{\overrightarrow{\rule{\transwidth}{0ex}}}}

  \author[1]{Andrea Benso}   
    \author[2,3]{Flavio D'Alessandro%\footnote{Corresponding author: dalessan@mat.uniroma1.it}
    }
      \author[2]{Paolo Papi} % \footnote{papi@mat.uniroma1.it}

 \affil[1]{Dipartimento di Matematica e Informatica ``U. Dini",
Universit\`a di Firenze, 
  50134 Firenze,  Italy}
  
 \affil[2]{Dipartimento di Matematica ``G. Castelnuovo",
 Sapienza Universit\`a di Roma,
  00185 Roma,  Italy}
  
 \affil[3]{Department of Mathematics, 
Bo\u gazi\c ci University,
  34342  Bebek, Istanbul, Turkey}

\begin{document} 

\maketitle

  \begin{abstract}  
    In this paper, we provide some conditions of algebraic and combinatorial flavour that make decidable the following problem:  
   given  a measure-once finite quantum automaton and a context-free grammar of finite index  or a simple matrix context-free grammar, 
  the set intersection of the language accepted by the automaton and the language generated by the grammar, is empty  or not.
 \end{abstract}

\noindent
{\bf Keywords:} {\em Quantum automata, Context-free languages, Algebraic gro\-ups, Matrix semigroups, Decidability}
\medskip

\noindent
{\bf AMS subject classifications:}  {\em 81P68, 68Q45, 03D05} % 03D05=Automata and formal grammars in connection with logical questions
\section{Introduction}
A fundamental issue in the mathematical theory of formal languages concerns the decidability of the {\em Emptiness Problem:}
given a model of computation $\cal Q$, is the language accepted by $\cal Q$, $L({\cal Q})$, equal to the empty set? A natural   % and straightforward 
generalization
of such a problem, called {\em Intersection Problem},  arises by relativizing it %  the previous problem
in the following way: % given a language $L$ taken from a predefinite family $\cal L$ and an instance of $\cal Q$, is the intersection of $L$ with $L()$ empty?
one considers  a predefined family $\cal L$ of (effectively defined) languages and asks whether the intersection of $L({\cal Q})$  with a language taken arbitrarily from $\cal L$ is empty.
Notice that, if $\cal L$ is the singleton set containing the free monoid of words over the input alphabet of $\cal Q$, one immediately obtains the Emptiness Problem.

In this paper,  by continuing an investigation of Bertoni, Choffrut et al. \cite{dlt2013,dlt2014}, 
we will study  the decidability of the {Intersection Problem} w.r.t.   quantum finite automata as the family  $\cal Q$  and
 context-free languages, together with some other strictly related languages, as  the family $\cal L$. %, together with some generalizations.
%
%
%
%In this paper,  by continuing an investigation of Bertoni, Choffrut et al. \cite{dlt2013,dlt2014}, we will study  the decidability of the {Intersection Problem} where  $\cal Q$ is the family of quantum finite automata and
%$\cal F$ is  the family of context-free languages and some other strictly related languages. %, together with some generalizations.
%
%
%
%
%In this paper,  we investigate the decidability of a problem for quantum automata introduced in \cite{dlt2013,dlt2014}, named {Intersection Problem}.
%In this paper, we investigate an issue of decidability  for quantum (finite) automata. % introduced in \cite{dlt2013,dlt2014}, named {\em Intersection Problem}.
%In this paper, we investigate the decidability of a problem for quantum automata introduced in \cite{dlt2013,dlt2014}, named {\em Intersection Problem}.
%
%
%
%
%Quantum automata were introduced at the beginning of 2000's in \cite{moore-crutchfield} as a new model of language recognizer 
%(see \cite{BMP03,MP21} for reference and updated texts). 
%
%
%
%
%
%
Quantum automata were introduced at the beginning of 2000's as a new model of language recognizer, that can be obtained by imposing the quantum paradigm on classical finite automata. Several models of quantum  automata have been proposed in the literature, which essentially differ in what quantum measurements are allowed
 (see \cite{AY2019,BMP03,Hin} for reference and updated texts). In this paper, we consider the model of a quantum automaton, called {\em ``measure-once"},  
 introduced by Moore and Crutchfield in  \cite{moore-crutchfield}. 
% Such a model is  the simplest one and provides the description of good-featured quantum devices of ``small size". In this regard, it is worth recalling that, 
% recently, in \cite{MP20,MP21}, a remarkable contribution of Mereghetti and Palano describes a method for the physical implementation of a suitable  class of quantum automata of such type. We describe a physical implementation of a quantum finite automaton that recognizes a well-known family of periodic languages.
% Such a model is  the simplest one and provides the description of good-featured quantum devices of ``small size". In this regard, it is worth recalling that, 
% recently, a remarkable contribution of Mereghetti and Palano describes a method for the physical implementation of  
% quantum automata of such type for the recognition of a well-known family of periodic languages \cite{MP20,MP21}.
Such a model is  the simplest  and one of the most studied in the literature since it 
provides the description of good-featured quantum devices of ``small size". In this regard, it is worth recalling that
a recent contribution of Mereghetti and Palano describes a method for the physical implementation of  
 measure-once quantum automata  for the recognition of a well-known family of periodic languages \cite{MP20,MP21}.

Numerous publications have investigated the decision properties of quantum automata.
%Numerous publications have compared the decision properties of quantum automata
%to those of other models and, in particular, to  probabilistic finite automata. 
% In this theoretical setting, we investigate the decidability of a problem for quantum automata introduced in \cite{dlt2013,dlt2014}, named {Intersection Problem}.
%
%Some undecidable problems 
%for probabilistic finite automata turn out to be  
%decidable for quantum finite automata.
%The result in  \cite{blondel} which triggered our investigation can be viewed as asserting that
%the intersection emptiness problem  of a language recognized by a finite 
%quantum automaton with the free monoid is   decidable.
%The present result concerns  the same problem 
%where instead of the free monoid,  more generally a language belonging to some classical 
%families of languages such as the context-free languages and the 
%bounded semilinear  languages is considered. 
%
The starting point of that investigation is a result of Blondel, Jeandel, Koiran, and Portier who showed in  \cite{blondel}  that, given a  
quantum automaton $\cal Q$ (in the sense of \cite{moore-crutchfield}), it is decidable whether the language recognized by $\cal Q$ has a non trivial  intersection  with the free monoid
generated by the input alphabet of $\cal Q$ (the Emptiness Problem, aforementioned). 
%
%
%In this paper, we investigate the decidability of a problem for quantum automata introduced in \cite{dlt2013,dlt2014}, named {\em Intersection Problem}.
%
%Quantum finite automata or simply quantum  automata were introduced at the beginning of the
%previous decade in \cite{moore-crutchfield} as a new model of language recognizer (cf. \cite{BMP03}). 
%Numerous publications have 
%ever since compared their decision properties
%to those of the older model of probabilistic finite automata. 
%Some undecidable problems 
%for probabilistic finite automata turn out to be  
%decidable for quantum finite automata.
%The result in  \cite{blondel} which triggered our investigation can be viewed as asserting that
%the intersection emptiness problem  of a language recognized by a finite 
%quantum automaton with the free monoid is   decidable.
%The present result concerns  the same problem 
%where instead of the free monoid,  more generally a language belonging to some classical 
%families of languages such as the context-free languages and the 
%bounded semilinear  languages is considered. 
%
%The starting point of that investigation is a result of Blondel, Jeandel, Koiran, and Portier which shows in  \cite{blondel}  that, given a finite
%(``measure-once" in the sense of \cite{moore-crutchfield}) quantum automaton $\cal Q$, it is decidable whether the language recognized by $\cal Q$ has a non trivial  intersection  with the free monoid
%generated by the input alphabet of $\cal Q$ ({\em Emptiness Problem}). 
%
%\bigskip
%{\bf ***}
%
It is worth remarking that these problems are  related to some computational issues concerning matrices. Precisely, a crucial step in the construction of the decision procedure
for both the Emptiness and Intersection Problem amounts to effectively compute the Zariski closure of a finitely generated group of matrices over $\Q$. 
A striking result proved by Derksen, Jeandel and Koiran in \cite{derksen} shows the existence of an appropriate algorithm for that task.
%For that task, an appropriate algorithm has been developed in \cite{derksen} by Derksen, Jeandel, and Koiran.  
% 
%\bigskip
%{\bf ***}

Before coming back to  the {Intersection Problem}, we find then convenient to recall some developments of the research on that aspect.
%
%
%
%\bigskip 
%{\bf ***}
%
Two % striking 
results have been obtained recently in connection with the algorithm  of \cite{derksen}. % mentioned above. 
In 2022
Nosan, Pouly, Schmitz, Shirmohammadi, and  Worrell,  following an alternative approach for such computation, 
have obtained  in \cite{npssw2022} a bound on the degree of the polynomials that define the Zariski closure of the set 
%and showing that the computation can be done in elementary time. 
and showed that the latter can be computed in elementary time. 
%
%
%\bigskip
%{\bf ***}

The second was obtained in 2023 by Hrushovski, Ouaknine, Pouly, and Worrell for the study of the effective computation of
polynomial invariants for affine programs \cite{hopw2023,hopw2023bis}. Precisely, it is  
exhibited an algorithm that, given a  finite set of square matrices of the same size with coefficients in $\Q$, computes the Zariski closure of the semigroup generated by the set.
% to compute the strongest polynomial invariants that hold at each location of a given a ne program (i.e., a program having only non-deterministic (as opposed to conditional) branching and all of whose assignments are given by a ne expressions). Our main tool is an algebraic result of inde- pendent interest: given a  nite set of rational square matrices of the same dimension, we show how to compute the Zariski closure of the semigroup that they generate.
%
%\bigskip
%
%{\bf ***}  
%
%

%In this context, it 
It should be pointed out that the decidability issues for matrix  semigroups
is a challenging problem. Even in small size, a well-known result by Paterson shows that the problem of freeness and that of
membership are both undecidable for matrix semigroups over $\Z$ of size $3$ (see \cite{CasNic}). In contrast, in 2017, a  surprising 
result of
Potapov and Semukhin \cite{potsam17} shows that the membership is decidable for non-singular matrix integer semigroups of size $2$.
%
%
%\bigskip
%
%{\bf ***} 
%

We now come back to the Intersection Problem and  give  a formal presentation of our work. 
% The free monoid generated by the finite alphabet
%$\Sigma$ is denoted by $\Sigma^*$.
%The elements of $\Sigma^*$ are \emph{words}.
%We consider all finite dimensional
%vector spaces as  provided with the Euclidian norm $||\cdot||$.
%
A \emph{finite quantum automaton} (in the sense of   \cite{moore-crutchfield}) is a quadruple

\begin{equation}\label{eq:quantum-automaton}
{\cal Q}=(s,\varphi,P,\lambda),
\end{equation} 
where $s\in \R^{n}$
is a row-vector of unit Euclidean norm,
$P$ is an  orthogonal projection matrix of size ${n}$,  $\varphi : \Sigma^* \rightarrow O_n$ is a morphism from the free monoid 
$ \Sigma^*$, generated by the input alphabet $\Sigma$ of $\cal Q$,  into the group $O_n$
of \emph{orthogonal} $n\times n$-matrices over $\R$ % $\R^{n\times n}$
and  the \emph{threshold}  $\lambda $  has value in $\R$.
%We recall that a real matrix $M$ is orthogonal if it is invertible and 
%its inverse equals its transpose: $M^{-1}=M^{T}$.
%We denote by $O_{n}$ the group of $n\times n$-orthogonal matrices.
%

We are mainly interested in effective properties
which require the quantum automaton
to be effectively given. 
We thus consider  the model of \emph{rational  quantum automaton}, i.e.,  where  all the coefficients 
%We thus consider the quantum automaton is \emph{rational} if  all the coefficients 
of the components of (\ref{eq:quantum-automaton}) and $\lambda$ % the automaton 
are rational numbers. % ,  i.e., $\varphi$ maps $ \Sigma^*$  into $\Q^{n\times n}$
% and $\lambda\in \Q$. 
%
%
%This hypothesis is not a  restriction 
%since all we use for the proofs is the fact 
%that the arithmetic operations and the comparison
%are effective in the field of rational numbers.
%
%
%The quantum automaton defined above is the {``measure once''}   \cite{moore-crutchfield}.
%The quantum automaton defined above is the {``measure once''} model introduced 
% by Moore and Crutchfield in 2000  \cite{moore-crutchfield}.
For a real threshold $\lambda$, the languages recognized by  $\cal Q$ with strict and nonstrict threshold $\lambda$ are respectively
 $$|{\cal Q}_{>}|=\{w\in  \Sigma^*\mid ||s \varphi(w) P|| ^2 >\lambda\}, \quad |{\cal Q}_{\geq}|=\{w\in  \Sigma^*\mid ||s \varphi(w) P||  ^2 \geq\lambda\},$$
 where $||\cdot||$ denotes the Euclidean norm of vectors.
   Let ${\cal L}$ be a  family of effectively defined languages. The  problem we tackle is the following:
%   We generalize the problem by considering 
%families of (effectively defined) languages ${\cal L}$ instead of the fixed language
%$\Sigma^*$. The  question we tackle is thus the following:
%

\medskip

\noindent
\underline{$(L, {\cal Q})$ Intersection}

\medskip

\noindent {\sc Input}:  a  language $L$ in a family ${\cal L}$ of 
languages  and a rational  quantum automaton
${\cal Q}$. 

\medskip

\noindent {\sc Question}:  does $L\cap |{\cal Q}_{>}|=\emptyset$ hold?
 
\medskip 

\noindent
As already pointed out, the decidability of this problem has been proven in  \cite{blondel}  in the specific case $L$  $= \Sigma^*$. % where $\cal L$ coincides with the free monoid $\Sigma^*$. 
%As already pointed out,   it has been proven in \cite{blondel} the 
%decidability of the problem in the specific case %  where 
%$L$  $= \Sigma^*$. % where $\cal L$ coincides with the free monoid $\Sigma^*$. 
This result contrasts with the corresponding problems (strict and non strict, respectively)  for probabilistic  automata. Indeed, for these automata, % where 
the Emptiness Problems w.r.t. $|{\cal Q}_{>}|$ and $|{\cal Q}_{\geq}|$  are  both undecidable 
\cite{Paz}, Thm 6.17 (see also \cite{BMT77} for the undecidability  of the property to be isolated for a cut point). 
Also it contrasts with the undecidability  of the Emptiness Problem of $|{\cal Q}_{\geq}|$ for quantum automata \cite{blondel}.
%Also it contrasts the undecidability of the non-strict version of the Emptiness Problem of $|{\cal Q}_{\geq}|$ for quantum automata \cite{blondel}.
%
%As already pointed out,   it has been proven in \cite{blondel} the 
%decidability of the problem in the specific case where $L$  $= \Sigma^*$. % where $\cal L$ coincides with the free monoid $\Sigma^*$. 
%This result  surprising contrasts the non-strict version of the Emptiness Problem of $|{\cal Q}_{\geq}|$ is undecidable. 
%It is worth remarking that these results  are in contrast with the corresponding ones  for probabilistic finite automata. Indeed for this class of automata % where 
%the above mentioned  problems   
%are  both undecidable (see \cite{Paz}, Thm 6.17). 
%As pointed out before, this problem has been studied by Blondel et al. in \cite{blondel}, where it has been proven its 
%decidability in the specific case where $\cal L$ coincides with the free monoid $\Sigma^*$. This result is surprising since it contrasts with the
%fact that 
% Blondel et al. in \cite{blondel} proved that the emptiness problem of 
%$|{\cal Q}_{>}|$ is decidable and the emptiness problem of 
%$|{\cal Q}_{\geq}|$ is undecidable. 
%It is worth remarking that these results  are in contrast with the corresponding ones  for probabilistic finite automata. Indeed for this class of automata % where 
%the above mentioned  problems   
%are  both undecidable (see \cite{Paz}, Thm 6.17). 
It is also proven in \cite{j02} that both the problems (strict and non strict, respectively) remain undecidable  for 
  the {``measure many''} model of quantum automata, a model not computationally equivalent to the {``measure once''},  introduced 
 by Kondacs and Watrous in \cite{kw}.
In order to describe the main contributions of the paper, it may be useful to recall some lines of the technique used in the proof of the 
Intersection Problem in \cite{dlt2014}. 
 For this purpose, following the   same strategy of \cite{blondel},  
we first observe that the Intersection Problem    is equivalent to the inclusion 

 \begin{equation}
\label{eq:l}
\varphi (L) \subseteq |{\cal Q}_{\leq}|.
\end{equation}
Since the function $$M \rightarrow ||s M P||  ^2,$$ 
-- where $M$ is an arbitrary matrix in $\R ^ {n\times n}$, and $s, P$ are components of (\ref{eq:quantum-automaton}) -- 
is continuous, it is sufficient to prove that, for all
matrices $M$  in the Euclidean closure $\Closure{\varphi(L)}$ of 
$\varphi(L)$,  the condition 
$||s M P|| ^2\leq \lambda$ holds  (see Section~\ref{sec: Effectiveness issues}).  
 Such condition is then effectively tested by  expressing the property (\ref{eq:l}) 
 in first-order logic of the field of reals, which, in turn, consists to effectively compute a representation of $\Closure{\varphi(L)}$ in term of a semialgebraic set 
(a more general property than
algebraic, which allows one to cope with the fact that algebraic sets are not closed w.r.t. the operation of product of sets).
 Afterwards, one  applies on the constructed formula 
the Tarski-Seidenberg quantifier elimination to verify whether (\ref{eq:l}) holds true or not. 

   % \medskip
 
It turns out that, in the case the language $L$ is context-free,
 the semialgebraicity of $\Closure{\varphi(L)}$, and its effective computation (as a semialgebraic set), % is based upon an algebro-combinatorial object that is
are reduced to those of an algebro-com\-bi\-na\-to\-rial object that is
{associated} with the grammar $G$ generating $L$, and with every variable $A$ of $G$: the {\em monoid of cycles of $A$.}
%{\em canonically associated} with the grammar $G$ spanning $L$, i.e. $L=L(G)$, and with every variable $A$ of $G$: the {\em monoid of cycles of $A$.}
Such monoid, denoted $M_A$,  corresponds (up to a technical detail) to the matrix image, under taking the morphism $\varphi$ of (\ref{eq:quantum-automaton}),  of the language of cycles in $G$
associated with $A$ (see Eq. (\ref{eq:crucial-monoid})). 

%It turns out that the effective computation of $\Closure{\varphi(L)}$, as a semialgebraic set, % is based upon an algebro-combinatorial object that is
%is reduced to an algebro-combinatorial object that is
%{\em canonically associated} with the grammar $G$ spanning $L$ and with every variable $A$ of $G$: the {\em monoid of cycles of $A$.}
%Such monoid, denoted $M_A$, corresponds to the matrix image, under taking the morphism $\varphi$ of (\ref{eq:quantum-automaton}),  of the language of cycles in $G$
%associated with $A$ (see Eq. (\ref{eq:crucial-monoid})). 

The reduction process mentioned above is made possible  by resorting to two ingredients: the fact that the Euclidean closure $\Closure{M_A}$ is an algebraic set
over the field $\R$ of real numbers,
 because of a deep result of Algebraic Geometry (see Theorem  \ref{th:topological-closure-of-monoid}) 
and because of a suitably defined effective structuring of the computations, i.e., the derivations,  of $G$ 
(Sections \ref{Sec:Structuring of a context-free language}, \ref{Sec:Structuring of FIN} and \ref{extens-BSL}).
In this theoretical setting, the work in \cite{dlt2014} and this work provide the following picture of the Intersection Problem:
%
%one can prove the following mathematical facts. 
\begin{itemize}
\item If $L$ is a context-free language, then $\Closure{\varphi(L)}$ is semialgebraic. Moreover, if, for every variable $A$ of $G$, 
$\Closure{M_A}$ is  effectively computable, then $\Closure{\varphi(L)}$  is so as well (Proposition \ref{main-prop-dlt2014}). 

%\item If $L$ is a context-free language, then $\Closure{\varphi(L)}$ is semialgebraic. Moreover, if, for every variable $A$ of $G$, 
%$\Closure{M_A}$ is  effectively computable, then $\Closure{\varphi(L)}$  is so as well (Proposition \ref{main-prop-dlt2014}). 

\item If, for every variable $A$ of $G$, $M_A$ is a regular submonoid, % finitely generated, 
then $\Closure{\varphi(L)}$ is  effectively computable. This is
guaranteed by applying Proposition~\ref{main-prop-dlt2014} and the results  \cite{derksen,hopw2023,npssw2022} mentioned above. 
This, in particular, covers the cases of context-free linear languages, bounded semi-linear languages and a new class of languages,  proposed in this paper, generated by 
{restricted matrix context-free grammars}  which non trivially extends the previous ones (Proposition~\ref{main-prop-gen-matr-gramm-bsl}).

\item The computation of $\Closure{\varphi(L)}$ is achieved for a subfamily of
  finite index context-free languages,  called {\em monoidal}, by considering the property of (algebraic) irreducibility on  the Zariski closure over $\R$ of a family of monoids of cycles {canonically associated} with  $G$ %  that yields $L$
 (Proposition \ref{prop-main-monoidal}).
% (Proposition \ref{prop-main-monoidal}, Corollary~\ref{cor-main-monoidal-1}, and the discussion below).

%\item The effective computation of $\Closure{\varphi(L)}$ is achieved by considering the property of (algebraic) irreducibility 
%on  the Zariski closure of some monoids of cycles {canonically associated} for a subclass of finite index grammars called monoidal%  that yields $L$
% (Proposition \ref{prop-main-monoidal}, Corollary~\ref{cor-main-monoidal-1}, and the discussion below).

%\item  In the case of finite index context-free languages,  if the Zariski closure of the monoids of cycles of the linear grammars of the lowest level of the  
%Ginsburg and Spanier decomposition of grammar generating the language,  are (algebraic) irreducible,
%then one can effectively computes the Euclidean closure $\Closure{\varphi(L)}$ of 
%$\varphi(L)$. %  (Corollary \ref{cor-main-monoidal-1}).

\end{itemize}
%The general case of context-free languages is still open and it would be even possible that  the problem is undecidable (cf Section \ref{sec: final comments}). 
%
%
%If $L$ is a context-free language, then $\Closure{\varphi(L)}$ is semialgebraic. Moreover, if, for every variable $A$ of $G$, 
%$\Closure{M_A}$ is  effectively computable, then $\Closure{\varphi(L)}$  is so as well (Proposition \ref{main-prop-dlt2014}). 
%
%Here we show that the (actually semi-)algebraicity of ${\cal A}$  still  holds
%when considering not only the free monoid but more generally 
%arbitrary context-free languages and bounded semilinear languages. 
%Unfortunately, its effective construction is only guaranteed under stricter 
%conditions such as the fact 
%that the language is context-free  linear or is bounded semilinear. 
%
%
% \bigskip
%
%
%
%
%\medskip 
The new results of the paper are the following. In Section \ref{extens-BSL}, 
we extend the main contribution of  \cite{dlt2014} to  a subfamily of {\em simple matrix languages.}
Simple matrix  languages have been introduced by Ibarra in \cite{ibarra-sml} as a  proper class of the deterministic linear-bounded acceptor languages,
defined by a restricted form of matrix context-free grammars (see \cite{ibarra-mcquillan}  for recent developments). They contain, in particular, context-free languages and inherit several
structural and decidable properties. 
  Precisely, we prove that, whenever the quantum automaton is rational, such extension (Proposition~\ref{main-prop-gen-matr-gramm-bsl}) 
 allows one, on one hand, to recover both the previous two cases -- i.e., the case of linear context-free languages and that of bounded semi-linear languages --   as immediate  
 corollaries. On the other hand, it shows  the decidability of the problem for  languages of exponential growth that are not context-free. 
  In the proof, a suitable decomposition of the derivations in these grammars is used. Such a decomposition is based upon some combinatorial tools proposed by Choffrut, Varricchio and the second author in \cite{tcsCDV2010} 
 for the study of the counting functions of rational relations. 
  In particular, this decomposition permits a refinement of the machinery developed in \cite{dlt2014} for the
 manipulation of the monoids of cycles of these grammars. 

In  Section \ref{subsec: the case of monoidal languages}, we investigate the Intersection Problem for languages defined by {\em finite index context-free grammars}. 
%A context-free grammar $G$ is said to be {\em of finite index} if 
These are grammars $G$ where each word $w$ in the language generated by $G$ is obtained by some derivation~$\delta$, whose index
is uniformly bounded, i.e., the number of variables in each sentential form of $\delta$ is bounded by an integer not depending on $w$. 
A result by   Ginsburg and Spanier \cite{GinSpan} (see also Nivat \cite{Niva}) provides a characterization of such languages in terms of composition of grammars:
precisely, each language of this type is generated by a grammar $G$ given by the composition
\begin{equation}\label{eq:intro-FI-gramm}
G = {\cal G}_1 \circ {\cal G}_2 \circ \cdots \circ {\cal G}_k,
\end{equation}
of  families  ${\cal G}_i, \ 1\leq i \leq k,\ $ of context-free linear  grammars. 
Here, we prove a statement of decidability for  a  subclass of these grammars called {\em monoidal}. 
 A finite-index grammar $G$     is  monoidal if the grammars of  (\ref{eq:intro-FI-gramm})  are minimal linear,  
 and the terminal productions of $G$   are of the form $X\rightarrow \varepsilon$   (see Definition \ref{defo:monoidal grammar}).
% 
% 
% 
% 
% 
% Informally speaking, a grammar $G$ of the form (\ref{eq:intro-FI-gramm})  
%    is  monoidal if
%the terminal productions of $G$   
%are of the form $X\rightarrow \varepsilon$ 
%and the remaining  productions are unit  ones or of the form
%$X\rightarrow uXv$, with $X$ being a variable of some grammar of ${\cal G}_i, \ 1\leq i\leq k$  (see Definition \ref{defo:monoidal grammar}).
%
%
%
%
%
%
%
%
%
%
%
%
%
%
% Informally speaking, a grammar $G=\langle V, \Sigma, P, S \rangle$ of (\ref{eq:intro-FI-gramm})  is  monoidal if
%the terminal productions of $G$   
%are of the form $X\rightarrow \varepsilon$ % ($X$ being a variable of $G$) 
%and the remaining ones are unit productions or of the form
%$X\rightarrow uXv$, with $u, v \in \Sigma^*$ (Definition \ref{defo:monoidal grammar}).
%
%
%
%
% A grammar $G=\langle V, \Sigma, P, S \rangle$  is said to be monoidal if
%the terminal productions of $G$ % , or equivalently those of ${\cal G}_i$ in  (\ref{eq:intro-FI-gramm}),  
%are of the form $X\rightarrow \varepsilon$ ($X$ being a variable of $G$) and the remaining {\em ``non trivial ones"} are of the form
%$X\rightarrow uXv$, with $u, v \in \Sigma^*$ (Definition \ref{defo:monoidal grammar}).

We prove that, regardless of the length $k$ of the composition  (\ref{eq:intro-FI-gramm}) for $G$,
if the Zariski closure over $\R$ of the monoids of cycles of the grammars of the lowest level  ${\cal G}_k$ of (\ref{eq:intro-FI-gramm}), are (algebraic) irreducible,
then one can compute the Euclidean closure $\Closure{\varphi(L)}$ of $\varphi(L)$  (Proposition \ref{prop-main-monoidal}),
%(Proposition \ref{prop-main-monoidal}, Corollary~\ref{cor-main-monoidal-1}),
thus making the Intersection Problem decidable % in this case 
(Corollaries~\ref{cor-main-monoidal-1} and~\ref{cor-main-monoidal-2}).
Extensions of such result to broader families of finite index context-free languages seem related to possible extensions of the algorithms
\cite{derksen,hopw2023,npssw2022}   (cf Section \ref{sec: final comments},  [$\alpha$]). 
The general case of context-free languages is open and it could be even possible that  the problem is undecidable (cf Section \ref{sec: final comments},  [$\beta$]).

The paper is structured in the following way. In Section \ref{sec: Preliminaries on semialgebraic sets} preliminaries on semialgebraic and irreducible algebraic sets are presented. 
In Section \ref{sec: Effectiveness issues} we recall an effective condition % of \cite{blondel} 
to which the decidability of the problem under discussion can be reduced. 
Section \ref{sec:context-free} provides %  is devoted to suitable 
combinatorial structurings of context-free languages in the general case and in the finite index case. 
Section~\ref{subsec:cf-lin} recalls some results % of \cite{dlt2014} 
relating context-free grammars and semialgebraic sets.
 In Section \ref {sec: main results} the main results of the paper are presented.
Section \ref{sec: final comments} contains some concluding remarks and open problems.
%Section \ref{Sec: Technical details and Examples} provides an example showing some basic points of the construction for monoidal grammars.
%Section \ref{Sec: Technical details and Examples} 
In the Appendix, % a technical remark on quantum automata and  
the construct of composition of grammars is given in detail. % on an example.

The results of this paper have been announced at RP 2024 \cite{bdp2024}. The present paper contains fully detailed proofs and examples of
applications and construction.

 \section{Preliminaries on semialgebraic sets} \label{sec: Preliminaries on semialgebraic sets}
 Let us give first the definition of quantum automaton (cf. \cite{dlt2013,blondel,moore-crutchfield}). 

  \begin{definition}\label{main-def-qa-inreal-form}
A quantum automaton  ${\cal Q}$ is a quadruple  $(s,\varphi, P,\lambda)$
where $s\in \R^{n}$
is a vector of unit norm,
%$P$ is a projection of $\R^{n}$, i.e., $P\in \{0,1\}^{n\times n}$ with $P^2=P$, and
%$P$ is an orthogonal projection of $\R^{n}$, and  
$P$ is an orthogonal matrix of size $n$ with   $P^2=P$, and
$\varphi$ is a morphism 
\begin{equation}\label{eq:main-morphism}
\varphi :   \Sigma^* \longrightarrow O_n,
\end{equation}
of the free monoid  $ \Sigma^*$  into the group $O_n$
of orthogonal $n\times n$-matrices in  $\R^{n\times n}$.
 \end{definition}
    \begin{remark}  
    Throughout the paper we will assume that  the linear representation of every quantum automaton (with possibly complex entries) is given,
     as in Definition \ref{main-def-qa-inreal-form}, by matrices over the field of real numbers.
This is a merely technical restriction since any  quantum automaton can be simulated by another one of the latter type. % (see the Appendix for details). 
We refer the reader to the Section ``Complex versus real entries" of
the paper  \cite{blondel}, where this issue is very clearly and thoroughly
explained.
      \end{remark}
The behaviour of ${\cal Q}$  heavily depends 
on the topological properties of the semigroup of matrices
$\varphi(\Sigma^*)$. This is why, before returning
to quantum automata, we first focus our attention
on these matrices for their own sake.

\subsection{Topology}

The following result is needed in the proof of the main theorem.
It will be applied in the particular case of orthogonal
matrices. 
%Though valid under weaker conditions,
%it will be considered in the particular case of orthogonal
%matrices. 
Given a subset $E$ of a finite dimensional vector space, we denote by 
$\closure(E)$ the topological closure 
for the topology induced by the Euclidean norm.
Given a $k$-tuple of matrices $(M_{1}, \ldots, M_{k})$, 
 denote by $f$  the $k$-ary product 
 $f(M_{1}, \ldots, M_{k})=M_{1} \cdots M_{k}$    % {\cal A}
 and extend the notation to subsets  ${\cal A}$ 
 of $k$-tuples of matrices by posing 
 $f({\cal A})=\{f(M_{1}, \ldots, M_{k})\mid (M_{1}, \ldots, M_{k}) \in {\cal A}\}$.
The following result will be applied in several instances of this paper.
It says that because we are dealing with compact subsets,  the two operators
of matrix multiplication and  the topological
closure   commute.  
\begin{theorem}{(\cite{dlt2014}, Theorem 1)}
\label{th:fundamental} 
Let $\cal C$ be  a compact subset of matrices  and let ${\cal A} \subseteq {\cal C}^k$ 
be a $k$-ary relation. Then we have $\Closure{ f({\cal A})}= 
f( \Closure{{\cal A}})$.

 \end{theorem}

 \noindent
Thus, if ${\cal A}$ is a   binary
 relation which is a direct product ${\cal A}_{1}\times {\cal A}_{2}$, 
we have $\Closure{{\cal A}_{1} {\cal A}_{2}}=f(\Closure{{\cal A}_{1}\times {\cal A}_{2}})$.
It is an elementary 
result of topology  that 
$\Closure{{\cal A}_{1}\times {\cal A}_{2}}=\Closure{{\cal A}_{1} }\times\Closure{{\cal A}_{2} }$ holds.
Because of 
$\Closure{{\cal A}_{1} {\cal A}_{2}}=f(\Closure{{\cal A}_{1}\times {\cal A}_{2}})
=f(\Closure{{\cal A}_{1}}\times \Closure{{\cal A}_{2}})
=\Closure{{\cal A}_{1} }\  \Closure{{\cal A}_{2} }$, 
we have the following %corollary
\begin{corollary}
\label{cor:closure-of-product}
The topological closure of the product of two sets of matrices 
included in a compact subspace is equal to the product of the topological
closures of the two sets.
 \end{corollary}

 \subsection{Algebraic and semialgebraic sets}
% \bigskip
%
%\noindent
%{\bf *** DARE bene  la defo di {\em "effectively algebraic set"}  ***}
%\bigskip
%
%\noindent
%
%
%
%
%%
Let us give first the definition of algebraic 
set over the field of real numbers (cf. \cite{BasuetAl,Ge,oni}). 
 \begin{definition}
A subset ${\cal A} \subseteq \R ^ {n}$ is   
{\em algebraic (over the field of real numbers)},  if  
${\cal A}$ is the zero set of an arbitrary subset $\cal P$ of polynomials
  of  $\R[x_{1}, \ldots, x_{n}]$, i.e.,
for every vector  ${\bf v} \in  \R ^ {n}$,

  \begin{equation}
  \label{eq:infinite-polynomials}
  {\bf v} \in {\cal A} \ \Longleftrightarrow \ \forall \ p\in {\cal P}: \ p({\bf v})=0.
  \end{equation}
%$$  v \in {\cal A} \ \Longleftrightarrow \ \forall \ p\in {\cal P}: \ p(v)=0.$$
%The set ${\cal A}$ is said to be  {\em effective algebraic} if the polynomials of $\cal P$ can be computed by an effective procedure  (algorithm).
 \end{definition}
Note that by the Hilbert's  basis theorem (see \cite{Ge}, Ch. 1, Sec. 1, Theorem~$1$), one may assume that the set  $\cal P$ % of the previous definition 
is finite.   
%Note that by the Hilbert basis theorem, one may assume that the set  $\cal P$ % of the previous definition 
%is finite.  
%
% 
%The equivalence of the two statements is an immediate consequence
%of the Hilbert finite basis Theorem. 
%Indeed, the theorem claims that given a 
%family  ${\cal P}$ there exists a finite
%subfamily $p_{1}, \ldots, p_{r}$ generating the same 
%ideal which implies in particular that for all $p\in {\cal P}$ there exist
%$q_{1}, \ldots, q_{r}$ with 
%%
%$$
% p= q_{1}p_{1} + \cdots + q_{r}p_{r}
%$$
%%
%Then $p_{j}(v)=0$ for $j=1, \ldots, r$ implies $p(v)=0$. 
Even more,  since we are dealing with algebraic sets over $\R$, $\cal P$ can be reduced to a singleton since (\ref{eq:infinite-polynomials}) can be assumed to be 
to the equation
 $$\displaystyle \sum _{p\in \cal P} p({\bf v})^2=0.$$
%
% $$\displaystyle \sum^n_{i=1} p_{j}(x)^2=0$$
 %
%As a trivial example, a singleton $\{v\}$ is algebraic since it is the unique
%solution of the equation $$\displaystyle \sum^n_{i=1} (x_{i}-v_{i})^2=0$$
%where $v_i,$ with $1\leq i\leq n,$ is the i-th component of the vector $v$. 

\noindent
One can check that
the family of  algebraic sets  is closed under finite unions
and intersections. 
However, it is not closed under  complement and projection.
 The following more general class of subsets enjoys 
extra closure properties and is thus more robust.
The equivalence of the two definitions
below is guaranteed by  Tarski-Seidenberg  quantifier elimination result (cf \cite{BasuetAl}, Ch. 2, Sec. 2.3 and Sec. 2.5, Cor. 2.95).

 \begin{definition} \label{def-semi-alg-set}
A subset ${\cal A} \subseteq \R ^ {n}$ is   
 {\em semialgebraic (over the field of real numbers)} if it satisfies one of the two
 equivalent conditions
 
 \medskip

\noindent {(i)} ${\cal A}$ is the set of vectors satisfying 
a finite Boolean
combination of predicates of the form $p(x_{1}, \ldots, x_{n})>0$ 
where $p\in \R[x_{1}, \ldots, x_{n}]$.

 \medskip

\noindent {(ii)} ${\cal A}$ is {\em first-order definable} in the theory of
 the structure whose domain are the reals 
and whose predicates are of the form 
$p(x_{1}, \ldots, x_{n})>0$ or $p(x_{1}, \ldots, x_{n})=0$ 
with $p\in \R[x_{1}, \ldots, x_{n}]$.
 
%The set ${\cal A}$ is said to be  {\em effective semialgebraic} if the formula defining ${\cal A}$ can be effectively computed. %  by an effective procedure  (algorithm).
 \end{definition}
%
%
%
%
%
%
%We now introduce the more general class of {\em semialgebraic sets} that enjoys 
%extra closure properties and is therefore more robust. To this purpose, we follow (\cite{BasuetAl}, Ch.2, Sec. 2.3).
% \begin{definition} \label{def-semi-alg-set}
% The family of {\em semialgebraic sets} of $\R ^ {n}$ is
% defined as the smallest family of sets in $\R ^ {n}$, that contains the algebraic sets as well as the sets defined by {\em polynomial inequalities}, i.e.,
%  sets of the form $$\{(x_{1}, \ldots, x_{n})\in \R ^ {n} : p(x_{1}, \ldots, x_{n}) > 0\},$$ for some polynomial $p\in \R[x_{1}, \ldots, x_{n}]$, 
%  and which is also closed under the boolean operations (complementation, finite unions, and finite intersections). 
%  \end{definition}
% 
% \begin{remark}\label{rmk-semi-alg}
% It is useful to point out that a subset of $\R ^ {n}$ is semialgebraic  if and only if it is first-order definable (by a free-quantifer formula) in the theory of  the structure whose domain is the field of the reals and whose atoms are of the form
% $(p=0)$ or $(p>0)$, where $p\in \R[x_{1}, \ldots, x_{n}]$. 
%  \end{remark}
 
\noindent
%    We now specify such definitions to square matrices. 
We now instantiate the definitions above to the setting of square matrices.
 \begin{definition}
 A set  ${\cal A} \subseteq \R ^ {n\times n}$  of  matrices 
is {\em algebraic} (resp., {\em semialgebraic}) if considered as a
set of vectors of dimension $n^2$, it is algebraic (resp., semialgebraic).
\end{definition}
 \medskip

 \noindent
 {\bf Notational convention.} {\em In  connection with the $(L, {\cal Q})$ Intersection Problem, we  will adopt the following terminology.
A  set ${\cal A}$ of  matrices  
will be called  {\em effective algebraic} (resp., {\em effective semialgebraic}) if the polynomials (resp., the formula) defining ${\cal A}$ can be algorithmically computed
from the input of the problem, i.e., an effectively defined  language $L$ and  a finite quantum automaton $\cal Q$.   }

 \medskip

 \noindent

  We now combine the notions
of zero sets and of topology.
In the following we rephrase two results of
 \cite{blondel}  by emphasizing the main features that serve our purpose. Given a
subset $E$ of a group, we denote by $\langle E \rangle$ and by $E^*$ 
the subgroup and the submonoid generated by $E$, respectively.
% (for the proof of Theorem \ref{th:semigroup-in-compact} and  Theorem \ref{th:topological-closure-of-monoid}, see the Appendix).
   
\begin{theorem}
\label{th:semigroup-in-compact}
Let $S\subseteq \R^{n\times n}$ be a set of orthogonal  matrices
and let $E$ be any subset  of $S$ satisfying
$\langle S \rangle = \langle E \rangle$.
Then  we have $\Closure{S^*}=\Closure{\langle E \rangle}$. In particular $\Closure{S^*}$ is a group. 
\end{theorem}
\begin{proof} 
 It is known that every compact subsemigroup of a
compact group is a subgroup $G$.
Now $S^*\subseteq \langle E \rangle$
implies $G=\Closure{S^*}\subseteq \Closure{\langle E \rangle}$
and $S\subseteq G$ implies
$\Closure{\langle E \rangle}\subseteq G$
and thus  $\Closure{S^*}= \Closure{\langle E \rangle}$.
 \qed\end{proof} 
%
%
%\bigskip
%
%\noindent
%{\bf *** EUCLIDEAN closure = ZARISKI closure (ONLY) for MONOID }  
%  
%  \noindent
%The main consequence of the next theorem
%is that the  topological closure 
%of a monoid of orthogonal matrices 
%is algebraic (\cite{blondel}, Theorem 3.1, see also  \cite{oni}, Ch. 3, Sec. 4). 
The main consequence of the next theorem
is that the  topological closure 
of a monoid of orthogonal matrices 
is algebraic (\cite{oni}, Ch. 3, Sec. 4, Theorem~$5$, \cite{dlt2014}, Theorem $7$).

\begin{theorem}
\label{th:topological-closure-of-monoid}
Let $E$ be a set of orthogonal  matrices. 
Then   $\Closure{\langle E \rangle}$
is a subgroup of orthogonal matrices
and it is the zero set of 
all polynomials $p[x_{1,1},\ldots, x_{n,n} ]$
satisfying the conditions
$$p(I)=0 \  \mbox{ and } p(eX)=p(X)  \  \mbox{ for all  } e\in  E$$
Furthermore, if the matrices in $E$ have rational coefficients,
the above condition may be restricted to polynomials with
coefficients in $\Q$.

\end{theorem}

\subsection{Closure properties of semialgebraic sets}\label{Sec:Closure properties}
%In this paragraph we will recall some closure properties of
%the class  of effective algebraic and semialgebraic matrices investigated in  \cite{dlt2014}.
%First, we recall that
%the family of  effective algebraic sets  is closed under finite unions
%and intersections. 
%
%
The  \emph{product} of two sets of matrices is defined as
\begin{equation}
\label{eq:product-sets-matrices}
{\cal A}_{1} {\cal A}_{2}=\{M_{1}M_{2} \mid  M_{1}\in {\cal A}_{1}, M_{2}\in  {\cal A}_{2}\}.
\end{equation}
We define the  \emph{sandwich}  operation denoted by  $\diamond$
whose first operand is a set of pairs of matrices
 ${\cal A}\subseteq \R^{n\times n} \times \R^{n\times n} $ 
and the second operand a set of matrices ${\cal B}\subseteq \R^{n\times n} $
by setting  
$$
{\cal A} \diamond {\cal B}= \{XYZ\mid (X,Z)\in {\cal A} \mbox{ and } Y\in {\cal B}\}
$$
% The next operation will be used. 
Given a  bijection 
\begin{equation}
\label{eq:entry-renaming}
\pi: \{(i,j) \mid  i, j\in \{1, \ldots, n\}\}\rightarrow \{(i,j) \mid  i, j\in \{1, \ldots, n\}\}
\end{equation}
and a matrix $M \in   \R ^ {n\times n}$ we define a new operation by denoting $\pi(M)$ as 
the matrix $\pi(M)_{i,j}=M_{\pi(i, j)}$. Extend this operation to subsets 
of matrices ${\cal A}$. Write $\pi({\cal A})$ to denote the set
of matrices $\pi(M)$ for all $M \in   {\cal A}$.
\\
 The last operation is the
\emph{sum} of square  matrices  $M_{1},  \ldots ,  M_{k}$
whose result  is the square block matrix 
\begin{equation}  
\label{eq:sum-of-matrices}
M_{1}\oplus \cdots \oplus M_{k}=              \left (\begin{array}{cccc} 
            M_{1} &   0 &  \cdots &   0\\ 
            0 &   M_{2}   & \cdots &  0\\
            \vdots     & \vdots & \vdots & \vdots \\ 
             0 &   0  &   0  & M_{k}
            \end{array}
\right )
\end{equation}
These notations extend to subsets of matrices in the 
natural way.
Here we assume that all $k$ matrices have the same size $n\times n$. 
Observe that if the matrices are orthogonal,
so is their sum. Such matrices form a subgroup of
orthogonal matrices of size $kn\times kn$.

Now we  recall some closure properties of
the class  of semialgebraic sets of  matrices (see \cite{dlt2014}, Sec. 2.4).
  \begin{proposition}
\label{pr:union-prod-phi} 
%Let $ {\cal A}_{1},   {\cal A}_{2}\subseteq   \R ^ {n\times n}$ be two 
% effective algebraic sets of matrices, and 
%let $\pi$ be a mapping as in (\ref{eq:entry-renaming}). Then 
%$ {\cal A}_{1}  \cup {\cal A}_{2}$, $ {\cal A}_{1}  \cap {\cal A}_{2}$, $\pi( {\cal A}_{1})$ and $ {\cal A}_{1}{\cal A}_{2}$,  are  effective algebraic.
 %Let $ {\cal A}_{1},   {\cal A}_{2}\subseteq   \R ^ {n\times n}$ be two 
% effective algebraic sets of matrices. Then  so are 
%$ {\cal A}_{1}  \cup {\cal A}_{2}$, $ {\cal A}_{1}  \cap {\cal A}_{2}$ and $ {\cal A}_{1}{\cal A}_{2}$.
Let $ {\cal A}_{1},   {\cal A}_{2}\subseteq   \R ^ {n\times n}$ be two 
 sets of matrices, and let $\pi$ be a one-to-one mapping as in (\ref{eq:entry-renaming}). 

\medskip

\noindent 1) If $ {\cal A}_{1}$ and $ {\cal A}_{2}$  are algebraic (resp., effective algebraic) so are $ {\cal A}_{1}  \cup {\cal A}_{2}, {\cal A}_{1}  \cap {\cal A}_{2}$, and $\pi( {\cal A}_{1})$.

\medskip

\noindent 2) If $ {\cal A}_{1}$ and $ {\cal A}_{2}$  are semialgebraic (resp., effective semialgebraic),  so are 
$ {\cal A}_{1} \cup  {\cal A}_{2}$,  $ {\cal A}_{1}  {\cal A}_{2}$ and $\pi( {\cal A}_{1})$.
\end{proposition}

\begin{proposition}
\label{pr:sandwich}  Let $ {\cal A}_{1}\subseteq  \R ^ {n\times n}\times \R ^ {n\times n}$   
and $   {\cal A}_{2}\subseteq   \R ^ {n\times n}$ be semialgebraic (resp., effective semialgebraic).
Then $ {\cal A}_{1} \diamond   {\cal A}_{2}$ is  semialgebraic (resp., effective semialgebraic).
\end{proposition}

  \begin{proposition}
\label{pr:blocks-of-product} 
If $ {\cal A}_{1},   \ldots, {\cal A}_{k}\subseteq   \R ^ {n\times n}$ are  algebraic (resp., effective
algebraic, semialgebraic, effective semialgebraic), then so is the set
$ {\cal A}_{1}\oplus   \cdots \oplus{\cal A}_{k}$.

\end{proposition}
 
 \begin{proposition}
\label{pr:product-of-blocks} 
Let $ {\cal A}$ be a semialgebraic (resp., effective semialgebraic) set
of $kn\times kn$  matrices of the form (\ref{eq:sum-of-matrices}).
The set 
$
\{X_{1}\cdots X_{k}\mid X_{1}\oplus \cdots \oplus X_{k} \in {\cal A}\}
$
is  semialgebraic (resp., effective semialgebraic).
\end{proposition}

\subsection{Algebraic Irreducible sets}\label{Sec:Irreducible sets}
Let us consider the topological space obtained by equipping $\R^m$ with the {\em Zariski topology}  (see \cite{Ge}, Ch. $1$). 
 Given a non-empty subset $\cal Z$ of $\R^m$, $\cal Z$ is said to be {\em irreducible} if $\cal Z$ cannot be written as the union 
of two closed, i.e. algebraic, subsets of $\cal Z$, distinct from $\emptyset$ and $\cal Z$ itself. 
It can be proven that a non-empty algebraic subset $\cal Z$ of $\R^m$ is irreducible  if and only if
its vanishing ideal is a prime ideal  (cf \cite{Ge}, Prop. 1.1.12). 
Such sets play a role in the theoretical developments presented later (see Sec.   \ref{subsec: the case of monoidal languages}). 
Therefore, we recall a definition and  some results. 
%
% \begin{proposition}
%\label{pr:cartesian-product-irr-ET-other-OP}     
%{\bf *** INCIPIT da rimettere a POSTO (DARE la defo di IRR in ambiente defo)}
%\medskip
%
%\noindent
%  \end{proposition}
%
%Let $\R^m$ be the topological space obtained with the {\em Zariski topology.} 
%In the sequel, we consider the group GL$_m(\R)$ of invertible $m\times m$ matrices over $\R$ equipped with the {\em Zariski topology.} 
%Given a non-empty subset $\cal Z$ of GL$_m(\R)$, $\cal Z$ is said to be {\em irreducible} if $\cal Z$ cannot be written as the union 
%of two closed, i.e. algebraic, subsets of $\cal Z$, distinct from $\emptyset$ and $\cal Z$ itself. 
%It can be proven that a non-empty algebraic subset $\cal Z$ of GL$_m(\R)$ is irreducible  if and only if
%its vanishing ideal is a prime ideal  (cf \cite{Ge}, Prop. 1.1.12). 
%Irreducible sets will play a role in this theoretical setting (cf Sec.   \ref{subsec: the case of monoidal languages}). To this purpose, 
%we recall some basic results. 
%\medskip
%
%\noindent
%{\bf *** INCIPIT da rimettere a POSTO}
%\medskip
%
%\noindent
The next notation is useful.
 \medskip 
       
       \noindent
       {\bf Notation.} {\em The Zariski closure of a set $\cal A$ will be denoted by $\overline{\cal A}$.  } % \overline
% \medskip 
%       
%       \noindent
        \begin{definition}
Given non-empty algebraic sets ${\cal V} \subseteq \R^n$ and ${\cal W}  \subseteq \R^m$, a map $f:  {\cal V} \rightarrow~{\cal W}$ is said to be  {\em regular}\footnote{
By following a well-established vocabulary in Automata Theory, the term {\em regular} will be also used in Sec. \ref{sec:context-free} to name the family of subsets of monoids 
defined by finite automata}
 (or a {\em morphism of algebraic sets}), 
if there exist $p_1, \dots , p_m \in \R[x_1, .\ldots , x_n]$ (where the $x_i$ are indeterminates) such that, for all ${\bf x} = (x_1, \ldots, x_n)$ in $\cal V$
\begin{equation}
\label{eq:reg-map}
f({\bf x})=(p_1({\bf x}), \dots , p_m({\bf x})).
\end{equation}
  \end{definition}
The following fact is well known (cf \cite{Ge}, Remark 1.3.2).
\begin{lemma}
\label{lemma: regular maps}
If $f: {\cal V} \rightarrow {\cal W}$ is a regular map and   $\cal V$ is irreducible,
  then the Zariski closure $\overline{f({\cal V})}$ of ${f({\cal V})}$ is also irreducible.
    \end{lemma}
%
%\medskip
%
%\noindent
%{\bf FACT 2.4.1}\;
%{\em If $f: {\cal V} \rightarrow {\cal W}$ is a regular map and   $\cal V$ is irreducible,
%  then the Zariski closure $\overline{f({\cal V})}$ of ${f({\cal V})}$ is also irreducible.
%}
%\medskip  
%
%\noindent
We will now consider irreducible algebraic sets of the group GL$_m(\R)$ of invertible $m\times m$ matrices over $\R$ equipped with the {\em Zariski topology.} 
The next closure properties can be readily checked.
  \begin{proposition}
\label{pr:cartesian-product-irr-ET-other-OP}
  The family of irreducible algebraic sets is closed with respect to the operations  (\ref{eq:entry-renaming}), (\ref{eq:sum-of-matrices}) and the (finite) Cartesian product of sets. %, the product of sets, and the sandwich operation $\diamond $.
 \end{proposition}
\begin{proof}
Let $\cal A$ be an irreducible algebraic  set. By  Proposition \ref{pr:union-prod-phi}, with  $\pi$ as in (\ref{eq:entry-renaming}), 
$\pi({\cal A})$ is algebraic so $\overline{(\pi({\cal A}))}=\pi({\cal A})$ and the claim comes from Lemma~\ref{lemma: regular maps} since $\pi$ is a regular map.
For the operation (\ref{eq:sum-of-matrices}) and Cartesian product, the fact that, given irreducible algebraic sets $ {\cal A}_{1},   \ldots, {\cal A}_{k}$, then so is the set
$ {\cal A}_{1}\oplus   \cdots \oplus{\cal A}_{k}$ (resp., $ {\cal A}_{1}\times   \cdots \times {\cal A}_{k}$) comes from \cite{Ge}, Prop. 1.3.8-(a).\qed\end{proof}
%
%The proof of the next lemma is easy and will be omitted.
%The next lemma is used below and its proof is almost immediate.
The proof of the next lemma is almost immediate.
   \begin{lemma}\label{main-prop-main-monoidal-PART C-sub 1} 
   Let $H_1, \ldots, H_m$ be subsets of an arbitrary monoid $\mbox{M}$, every one of which contains the identity $1$ of M. Then
   $(H_1 \cup \ldots \cup  H_m)^* = (H_1\cdots H_m)^*$.
   \end{lemma}
   
   \begin{proof} 
   We prove the claim for $m=2$, i.e., $(A \cup B)^* = (AB)^*$ since the general case is  treated similarly. 
   The inclusion $(A \cup B)^* \supseteq (AB)^*$ is obvious. For the reverse inclusion, assume
   $m=m_1\cdots m_\ell, \ m_i\in A\cup B, \ 1\leq i \leq \ell$. Then $m$ can be written as
   $m =a_1b_1\cdots a_\ell b_\ell$, where, for every $i=1,\ldots, \ell$
   $a_i=m_i$ and $b_i=1$ if $m_i\in A$ while $a_i=1$ and $b_i=m_i$ if $m_i\in B$. 
   \qed   \end{proof}
   \noindent
        We now introduce an useful notation.
        \medskip 
       
       \noindent
       {\bf Notation.} {\em Given a subset \ $\cal A$\ of the group  $O_{m}$ of orthogonal matrices (over $\R$) of order $m\geq 1$, we write
 ${\cal A} \in ({\cal H})$ if
 \begin{equation}
\label{eq:(H)}
{\cal A} = \bigcup _{i=1} ^k {\cal A}_i, 
 \end{equation}
where, for every $i=1,\ldots,k$, ${\cal A}_i$ is an irreducible, effective algebraic set that contains the identity matrix $I$.}
%        if $\cal A$ is a finite union of irreducible, effective algebraic sets, every one of which
%       contains the identity matrix.}%  $I$ of $O_{m}$.}

   \begin{proposition}
\label{pr:Psi-Ir} 
Let $\cal A$ be a  subset of $O_{m}$ with ${\cal A}\in ({\cal H})$ and let $\Psi$ be  a regular map
%\begin{equation}
%\label{eq:Psi}
$$\Psi:  O_{m}  \;\; \displaystyle\mathop{\longrightarrow} \;\; O_{n}.$$
% \end{equation}
%be  a regular map such that $\Psi ({\cal A})$ contains the identity matrix $I$ of $O_n$.  
If, for every $i=1,\ldots,k$, $\Psi({\cal A}_i)$ contains $I$, 
then $\overline{\Psi({\cal A})^*} \in ({\cal H})$.
% In particular $\Psi({\cal A)}$ is effective algebraic.
%Then $\overline{\Psi({\cal A})^*}$ is a a  subset of $O_{n}$ with $\overline{\Psi({\cal A})^*}\in ({\cal H})$.
  \end{proposition}

%\noindent
%        We now introduce an useful notation.
%        \medskip 
%       
%       \noindent
%       {\bf Notation.} {\em Given a subset $\cal A$ of the group  $O_{m}$, of orthogonal matrices (over $\R$) of order $m\geq 1$, we write
% %       \begin{equation}
%% \label{defo:monoidalH}
%$${\cal A} \in ({\cal H})$$
%%\end{equation}
%       if $\cal A$ is a finite union of irreducible, effective algebraic sets, every one of which
%       contains the identity matrix.}%  $I$ of $O_{m}$.}
%
%
% 
%   \begin{proposition}
%\label{pr:Psi-Ir} 
%Let $\cal A$ be a  subset of $O_{nk}$ with ${\cal A}\in ({\cal H})$ and let 
%%\begin{equation}
%%\label{eq:Psi}
%$$\Psi:  O_{nk}  \;\; \displaystyle\mathop{\longrightarrow} \;\; O_{n}$$
%% \end{equation}
%be  a regular map such that $\Psi ({\cal A})$ contains the identity matrix $I$ of $O_n$.  
%Then $\overline{\Psi({\cal A})^*} \in ({\cal H})$.
%%Then $\overline{\Psi({\cal A})^*}$ is a a  subset of $O_{n}$ with $\overline{\Psi({\cal A})^*}\in ({\cal H})$.
%  \end{proposition}
%
%
%
%
%
%
%
%
% OLD VERSION PREVIOUS PROP
%
%  \begin{proposition}
%\label{pr:Psi-Ir} 
%Let $\cal A$ be an an irreducible, effective algebraic  set of $O_{nk}$, containing the identity matrix 
% and let 
%%\begin{equation}
%%\label{eq:Psi}
%$$\Psi:  O_{nk}  \;\; \displaystyle\mathop{\longrightarrow} \;\; O_{n}$$
%% \end{equation}
%be  a regular map such that $\Psi ({\cal A})$ contains the identity matrix $I$ of $O_n$.  
%Then $\overline{\Psi({\cal A})^*}$ is a finite union of irreducible, effective algebraic sets, every one of which
%       contains $I$.  
%  \end{proposition}

  \begin{proof}
  Let ${\cal A} = \bigcup _{i=1} ^k {\cal A}_i$ be a set as in (\ref{eq:(H)}).  % where, for every $i=1,\ldots,k$, ${\cal A}_i$ is an irreducible, effective algebraic set that contains $I$.
  We may suppose that $k=2$, i.e. ${\cal A} = {\cal A}_1\cup {\cal A}_2$, since the general case can be treated similarly. 
  Hence $\Psi({\cal A}) = \Psi({\cal A}_1\cup {\cal A}_2)$ so, by Lemma \ref{main-prop-main-monoidal-PART C-sub 1}, 
$$\Psi({\cal A})^* = (\Psi({\cal A}_1)\Psi({\cal A}_2))^*.$$

Let $H_1 =   \overline{\Psi({\cal A}_1)\Psi({\cal A}_2)}$. 
By Lemma \ref{lemma: regular maps}, $\overline{\Psi({\cal A}_i)}$ \  $(i=1, 2),$ 
 is an algebraic irreducible set. Further, by hypothesis, $\overline{\Psi({\cal A}_i)}$ contains $I$.
%   {\bf *** Point to FIX ***}
 Following \cite{derksen}, Sec. 3.1, by using {\em Gr\"obner bases techniques}, since ${\cal A}_i$ % \  (i=1, 2),$ 
is effective algebraic and $\Psi$ is a regular map, then one can effectively compute $\overline{\Psi({\cal A}_i)}$. %\  (i=1, 2)$. 
%
%
%  
%  Let $H_1 = \overline{\Psi ({\cal A})}$. 
%By Lemma \ref{lemma: regular maps}, $H_1$ is an algebraic irreducible set. Moreover, by hypothesis, $I\in H_1$.
%Following \cite{derksen}, Sec. 3.1, by using {\em Gr\"obner bases techniques}, since $\cal A$ is effective algebraic, then one
%can effectively compute $H_1$. 

\noindent
Now we observe that % , for arbitrary subsets ${\cal A}_1, {\cal A}_2$  of $O_n$
\begin{equation}\label{eq:main-example2-(pr:Psi-Ir)}
\overline{\overline{{\Psi(\cal A}_1)} \cdot \overline{\Psi({\cal A}_2})}\  = \ \overline{\Psi({\cal A}_1)\cdot\Psi({\cal A}_2)}.
\end{equation}
Indeed, taking the map m$: O_n \times O_n \rightarrow O_n$ defined, for every $X, Y\in O_n$, as m$(X,Y):= X\cdot Y$,
one has that m is a regular and thus continuous map % with respect to the Zariski topology on $O_{n}\times O_n$ and $O_{n}$,
(w.r.t. $O_{n}$ endowed with the Zariski topology and
$O_n \times O_n$ with the induced product topology). 
This implies, for arbitrary sets ${\cal X}_1, {\cal X}_2,$   % (by the same argument used in (\ref{eq:main-example1})) 
that $\overline{\mbox{m}(\overline{{\cal X}_1}, \overline{{\cal X}_2})} = \overline{\mbox{m}({\cal X}_1, {\cal X}_2)},$ and thus (\ref{eq:main-example2-(pr:Psi-Ir)}).
% Let $H_2 = \overline {H_1 \cdot H_1}$. 

\noindent
Let $H_2 = \overline {H_1 \cdot H_1}$. Since $I\in H_1$ then one has $I\in H_2$.
%Now we observe that, for arbitrary subsets ${\cal A}_1, {\cal A}_2$  of $O_n$
%\begin{equation}\label{eq:main-example2-(pr:Psi-Ir)}
%\overline{\overline{{\cal A}_1} \cdot \overline{{\cal A}_2}}\  = \ \overline{{\cal A}_1 \cdot {\cal A}_2}.
%\end{equation}
%Indeed, taking the map m$: O_n \times O_n \rightarrow O_n$ defined, for every $X, Y\in O_n$, as m$(X,Y):= X\cdot Y$,
%one has that m is a regular and thus continuous map % with respect to the Zariski topology on $O_{n}\times O_n$ and $O_{n}$,
%(w.r.t. $O_{n}$ endowed with the Zariski topology and
%$O_n \times O_n$ with the induced product topology). 
%This implies  % (by the same argument used in (\ref{eq:main-example1})) 
%that
%$\overline{\mbox{m}(\overline{{\cal A}_1}, \overline{{\cal A}_2})} = \overline{\mbox{m}({\cal A}_1, {\cal A}_2)}.$
%% Let $H_2 = \overline {H_1 \cdot H_1}$. 
By the very same argument used for $H_1$ one has that
$H_2$ is an irreducible effective algebraic set such that $H_2 = \overline{\Psi ({\cal A})^ 2}$.% $H_2 = \overline{(\Psi (G))^2}$.

\noindent
Finally, let us define recursively the family  of sets of $O_n$ $\{H_i\}_{i\geq 1},$ where $H_0= \{I\}$ and, for every $i\geq 1$, $$H_i = \overline{H_{i-1} \cdot H_1}.$$ 
By proceeding as in  the previous cases, one has that, for every $i\geq 3$, $H_i$ is an effective algebraic irreducible set such that 
%$H_i = \overline{\Psi ({\cal A})^i}$
$H_i = \overline{(\Psi({\cal A}_1)\Psi({\cal A}_2))^i}$. 
and containing $I$. Moreover,
  since, for each $i\geq 1$, $I\in H_i$, then one has the following ascending chain
\begin{equation}
\label{eq:-asc-chain}
H_1\;  \subseteq\;H_2\;  \subseteq\;  \cdots \; H_i\;  \subseteq\; \cdots 
\end{equation}
which, by the irreducibility of all of its terms, terminates, i.e., there exists some $d\in \N$ such that % (cf Remark~\ref{rmk1})
\begin{equation}
\label{eq:-eff-comp}
H_{d+1} \ = \  \overline{(\Psi({\cal A}_1)\Psi({\cal A}_2))^{d \ + \ 	1} } \ =\    \overline{(\Psi({\cal A}_1)\Psi({\cal A}_2))^d} \ = \ H_d.
%H_{m+1} \ = \  \overline{\Psi({{\cal A}})^{m \ + \ 	1} } \ =\    \overline{\Psi({{\cal A}})^m} \ = \ H_m.
%H_{m+1} \ = \ ( \overline{\Psi({G}))^{m \ + \ 	1} } \ =\   ( \overline{\Psi({G}))^m} \ = \ H_m,
\end{equation}
%   because the general linear group GL$(\R)_n$  has finite dimension 
   
   Finally let ${\cal X} = \bigcup _{i\geq 1}\ H_i$. By (\ref{eq:-eff-comp}), $\overline{\cal X} = H_1 \cup \cdots \cup H_d$. 
   On the other hand, since, for every $i\geq 1$, $H_i = \overline{(\Psi({\cal A}_1)\Psi({\cal A}_2))^i}$, then it is readily checked that
   $$\overline{\cal X} = H_1 \cup \cdots \cup H_d = \overline{\Psi ({\cal A})^*}.$$
   This proves the claim.
  \qed  \end{proof}

   \begin{remark}\label{rmk1}
      The integer $d$ of  (\ref{eq:-eff-comp}) can be bounded. Indeed, by \cite{Ge},
   Proposition 1.2.20 , assuming that   ${\cal V}, {\cal W}  \subseteq \R^{n}$ are algebraic sets and  ${\cal V}$ is irreducible,
    if  ${\cal V}  \subset {\cal W}$ then  dim$({\cal V})<$dim$({\cal W})$.    
     Since all the sets of (\ref{eq:-asc-chain}) are irreducible algebraic subsets of GL$_n(\R)$, which has dimension $n^2$, one has  $d\leq n^2$.
    \end{remark}
%   Observe that the integer $m$ of  (\ref{eq:-eff-comp}) can be bounded. Indeed, recall that, by \cite{Ge},
%   Proposition 1.2.20 , if   ${\cal V}, {\cal W}  \subseteq \R^{m}$ are algebraic sets and  ${\cal V}$ is irreducible, then  dim$({\cal V})<$dim$({\cal W})$,
%   if  ${\cal V}  \subset {\cal W}$.    % Recall now that the dimension of GL$(\R)_n$ is less or equal than $n^2$ (cf \cite{Ge}, Definition 1.2.15). 
%    Since all the sets of (\ref{eq:-asc-chain}) are irreducible algebraic subset of GL$_n(\R)$, which has dimension $n^2$, one has  $m\leq n^2$.
%    \end{remark}
%Propositions  \ref{pr:basic-irrid}  and \ref{pr:Psi-Ir} straightforwardly imply the following.
%
% \begin{corollary}
%\label{cor:ir-sets}
%{\bf  *** da SISTEMARE}
%   \end{corollary}
 
% \subsection{Effectiveness issues}Intersection
 \section{Effectiveness issues}\label{sec: Effectiveness issues}

 We now return to the $(L, {\cal Q})$ Intersection problem
as defined in the Introduction.  The main bulk of this decidability result is based upon the following
proposition that is obtained by proceeding as in \cite{blondel}. For the sake of completeness, we recall the proof.

 \begin{proposition}{(\cite{dlt2014}, Proposition 1)}
\label{pr:meta-resuositionlt}
Let ${\cal Q}$   be a rational quantum automaton. Let $L\subseteq \Sigma^*$ 
be a formal language such that   the set $\Closure{\varphi(L)}$ is effective
semialgebraic.  It is recursively decidable 
whether or not  
$L\cap |{\cal Q}_{>}|=\emptyset$ holds.
\end{proposition}
%Checking (ii) then amounts to deciding whether a first-order sentence of the language of ordered fields is true in the field of real numbers. It has been known since Tarski that this can be done algorithmically (more efficient algorithms and further references can be found in Basu et al. (1996) or Renegar (1992)).

\begin{proof}
%
%
% THE SUBSEQUENT PART HAS BEEN CHANGED 22.8.2024
%
%
% 
%
%Equivalently we prove the inclusion $L\subseteq |{\cal Q}_{\leq}|$. 
%We run in parallel two semialgorithms.
%The first one verifies the noninclusion by enumerating the words 
%$w\in L$ and testing if 
%$||s \varphi(w) P||>\lambda$ holds.
%%
%The second semialgorithm considers
%an effectively computable polynomial $\phi (X)$ 
%which defines 
%$\Closure{\varphi(L)}$ and verifies whether
%the sentence
%%
%$$
%\Psi\  \equiv\  \forall X: \phi (X) \Rightarrow
%sXP\leq \lambda
%$$
%%
%holds, which can be achieved by Tarski Seidenberg elimination 
%result.
%%
%If the inclusion $L\subseteq |{\cal Q}_{\leq}|$ 
%holds then the first
%semialgorithm cannot answer ``yes'' and the 
%second semialgorithm will 
%answer ``yes''. Now assume that the inclusion does not
%hold. Then the first
%semialgorithm will eventually answer ``yes'', while the second semialgorithm 
%cannot answer ``yes'' since this would imply that $\Psi$ holds and thus
%$$\forall X: X\in \Closure{\varphi(L)} \Rightarrow  ||sXP|| \leq \lambda, $$
% a contradiction.
% \qed\end{proof}
%%
%
%
%
%
%
%
%
Equivalently we prove the inclusion $L\subseteq |{\cal Q}_{\leq}|$, which is equivalent to
$$
 \forall \  w\in L \Rightarrow f(w)\ =\ ||s \varphi(w) P|| ^2 \ \leq \ \lambda.
$$
By the continuity of the map $f$, the last is equivalent to
\begin{equation}\label{pr:meta-resuositionlt-formula(1)}
 \forall \ X \in \Closure{\varphi(L)} \  \Rightarrow\ || sXP|| ^2 \ \leq\  \lambda.
\end{equation}
By hypothesis, there exists an effectively constructible formula $\Phi(X)$ that defines $\Closure{\varphi(L)}$ as a semialgebraic set (see Definition \ref{def-semi-alg-set}).
Hence (\ref{pr:meta-resuositionlt-formula(1)}) can be written as 
$$ \Upsilon\;  \equiv\;  \forall  \ X: \Phi(X) \  \Rightarrow\ || sXP||  ^2\ \leq\  \lambda.
$$
This implies that $L\subseteq |{\cal Q}_{\leq}|$ if and only if $\Upsilon$ holds true. 
Finally, one verifies whether 
$\Upsilon$ holds, which can be achieved by applying the Tarski Seidenberg elimination 
result.
  \qed\end{proof}
Proposition \ref{pr:meta-resuositionlt}
 shows that the decidability of the problem
relies upon the computation of $\Closure{\varphi(L)}$. In the sequel, 
we will investigate the case of context-free
languages. In order to deal with this case, we will use  two main tools: first, an algorithm by Derksen et al. \cite{derksen},  already mentioned in the Introduction, that computes
the Zariski closure of a finitely generated group of matrices over a field; secondly,
a suitable combinatorial structuring of a context-free
language based upon some special structures named cycles. The last will be presented in the next section.  
 
\section{Preliminaries on context-free languages}
\label{sec:context-free}
For the sake of self-containment 
and in order to fix notation, we just recall few essential terms  
 of context-free languages and regular sets of monoids, that can be found in all 
introductory textbooks on theoretical computer science (see, for instance, \cite{Ber79,Gins,Hu}).
%
%For the sake of self-containment 
%and in order to fix notation, we recall the basic properties 
%and notions concerning the family
%of context-free languages which can be found in all 
%introductory textbooks on theoretical computer science (see, for instance, \cite{Ber79,Hu}).
%
%

Given  a monoid $M,$ a subset of $M$ is said to be {\em regular} if it is finite or it is obtained from finite sets
by using finitely many times the {\em regular operations}, i.e., the union of two sets, the product $(\cdot)$ of two sets, and the
star $(^*)$, this last operation associating with a set $\cal X$ the submonoid ${\cal X}^*$ generated by $\cal X$. 
It is worth noticing that the concept of regular set can be equivalently given in terms of $M$-automaton. We will use this fact in the sequel (cf  Lemma \ref{main-lm-lin:dlt2014},
 Proposition \ref{main-prop-gen-matr-gramm-bsl}).
%We will recall this equivalence in the appropriate points (cf  Lemma \ref{main-lm-lin:dlt2014},
% Proposition \ref{main-prop-gen-matr-gramm-bsl}).

A  \emph{context-free grammar} $G$ is a quadruple $\langle V, \Sigma, P, S \rangle$ 
 where $\Sigma$  is the  alphabet of   \emph{terminal symbols},
$V$ is the set of \emph{variables} or \emph{nonterminal symbols}, $P$ is the set of 
\emph{rules} or \emph{productions}, and $S$ is the \emph{axiom} of the grammar.  
 A word over the alphabet $\Sigma$ is called {\em terminal}.
As usual, the nonterminal symbols are denoted by uppercase 
letters as, for instance, $A$, $B$. An arbitrary rule of the grammar is written
as $A\rightarrow \alpha$, where $\alpha$ is a finite sequence of variables and terminals.  
 The 
\emph{derivation} relation of $G$ is denoted by $ \displaystyle\mathop{\Rightarrow}^{*}$.
The {\em language generated by $G$} is the subset of terminal words  $ \{w\in \Sigma ^* : S\displaystyle\mathop{\Rightarrow}^{*} w  \}$ denoted by $L(G)$.
The family of context-free languages is denoted by CFL. 

Following \cite{Ber79}, Sec. VII.5, we say that a  context-free language $L$ is   {\em of  index $k$}, % with $k\in \N$,  % or {\em quasi-rational} 
if there exist a context-free grammar $G$ with $L=L(G)$  and an integer $k\in \N$ such that the following property holds:
for every word $w\in L(G)$, there exists some derivation  $\delta=(S\displaystyle\mathop{\Rightarrow}^{*} w)$  such that the number of occurrences of variables in each sentential form of 
$\delta$ does not exceed $k$.
The family of context-free languages of index $k$ will be denoted by $Ind(k)$ and  $\FIN$ will denote the family of all the languages {\em of finite index}, that is, $\bigcup _{k\in \N} Ind(k)$. 
One immediately sees that the most simple languages in such a family  are those of $LIN$,  % \emph{linear} ones, i.e. those 
the {\em linear} languages, i.e. those generated by  grammars 
 where the right hand side $\alpha$ of every rule $A\rightarrow \alpha$, contains at most one occurrence of nonterminal symbols,
that is, $\alpha \in \Sigma^{*}  \cup \Sigma^{*} V \Sigma^{*}$.
 A well-known result proven by Salomaa in \cite{s} shows that Dyck languages cannot be generated by finite index grammars, thus implying 
$\FIN \subset \ ${CFL}.
 The importance of the family $\FIN$ is due to the fact that, by a well-known result by Baron and Kuich \cite{baronKuich} %  (see also \cite{CarpiDal2018}), 
 (up to a technical restriction) the unambiguous context-free languages whose characteristic series in commutative variables are rational, 
are exactly those in $\FIN$.   
% This also justifies the term ``quasi-rational" for such languages.
For this reason, such languages are also called {\em quasi-rational} (see Section \ref{Sec:Structuring of FIN}).

     \subsection{Structuring of {CFL}}\label{Sec:Structuring of a context-free language}
 We now find convenient to recall a combinatorial structuring of an arbitrary context-free language (see \cite{dlt2014}, Sec. 3). Such decomposition follows a recursive scheme proposed by Incitti in
 \cite{I} based upon the concept defined below.
%
% ``contexts'' or ``cycles'' 
%as defined by the following.
 \begin{definition}
\label{de:contexts}
With each nonterminal symbol $A\in V$ associate the subset of $\Sigma ^* 
\times \Sigma ^*$  defined as  
$$
\G_A = \{(\alpha, \beta) \in \Sigma ^* 
\times \Sigma ^*: A \displaystyle\mathop{\Rightarrow}^{*} \alpha A \beta \}.
$$
\end{definition}
$\G_A$ will be called the {\em set of cycles of $A$}.\ \footnote{According to the terminology of \cite{Gins},  
a {\em cycle} is a derivation of the form $A \displaystyle\mathop{\Rightarrow}^{*} \alpha  A \beta$. 
With a minor abuse of language, we will call cycle  the pair of contexts $(\alpha,\beta)$ also} 

%\begin{definition}
%\label{de:contexts}
%With each nonterminal symbol $A\in V$ associate the subset of $\Sigma ^* 
%\times \Sigma ^*$ of its
%\emph{terminal contexts} or \emph{cycles}  defined as  
%%
%$$
%\G_A = \{(\alpha, \beta) \in \Sigma ^* 
%\times \Sigma ^*: A \displaystyle\mathop{\Rightarrow}^{*} \alpha A \beta \}.
%$$
%%
%\end{definition}

\noindent
It is convenient to define the sandwich operation (cf Sec. \ref{Sec:Closure properties}) also for  languages in the following way.  If $C_A$ is the set above and  $L'$ an arbitrary language, 
 we define the subset of $\Sigma ^*$
  \begin{equation}
   \label{eq:sandwich-op}
   C_A \diamond L' = \{uwv\mid (u,v)\in {C_A} \mbox{ and } w\in {L'}\}.
   \end{equation}
As the construction of the structuring mentioned above proceeds by induction
on the number of nonterminal symbols, we need to show how to 
recombine a grammar from simpler ones
obtained by choosing an arbitrary non-axiom symbol
as the new axiom and by removing all the rules 
involving $S$. This is the reason for introducing
the next notation.

 \begin{definition}
 \label{de:v-prime}
 Let $G = \langle V, \Sigma, P, S \rangle$ be a  context-free grammar. 
Set $V' = V \setminus \{S\}$.
 
For every $A\in V'$, define the context-free      
grammar $G_{A} = \langle V', \Sigma, P_{A}, A \rangle$ where
the set $P_{A}$  consists of all the rules $B \rightarrow \gamma$ of $G$  of the form
$$
\quad B \in V', \quad \gamma \in (V' \cup \Sigma)^*
$$
and denote by $L_A$ the language of all terminal words 
generated by the grammar $G_A$.
 \end{definition}
The next definition introduces the language of terminal words
obtained in a derivation where   
$S$ occurs at the start only.

  \begin{definition}
 \label{de:L-prime}
Let  $L'(G)$  denote the set of all the words 
of $\Sigma^*$ which admit a derivation
   \begin{equation}
   \label{eq:derivation}
   S \Rightarrow \gamma_{1} \Rightarrow \cdots  \Rightarrow \gamma_\ell  \Rightarrow w
   \end{equation}
where, for every $i=1, \ldots, \ell$,   $\gamma _i\in (V' \cup \Sigma)^*$.

%If no ambiguity arises, in the sequel, the language  $L'(G)$  is simply denoted $L'$.
 \end{definition}

The language $L'(G)$ can be easily expressed in terms 
of the languages $L_{A}$ for all $A\in V'$. Indeed,
consider the set of all rules of the grammar $G$ of the form
 \begin{equation}
  \label{eq:beta}
S\rightarrow \beta, \quad \beta \in (V' \cup \Sigma)^*
\end{equation}

Factorize every such $\beta$ as
 \begin{equation} 
 \label{eqfla:rob}
 \beta =w_{1}A_{1}w_{2}A_{2} \cdots w_{\ell}A_{\ell} w_{j_{\ell+1}}
 \end{equation}
where
$w_{1},  \ldots, w_{\ell+1}\in \Sigma^*$ and $A_{1}, A_{2}, 
\ldots, A_{\ell} \in V'$.
 %
%Next lemma easily follows from the previous definitions.  

%\begin{lemma}
%\label{le:L'}
% With the notation of (\ref{eqfla:rob}), 
%the language $L'(G)$ is  the finite
% union of the languages 
%%
%$$
%w_{1}L_{A_{1}}w_{2}L_{A_{2}} \cdots w_{\ell}L_{A_{\ell}} w_{j_{\ell+1}}
%$$
%% 
%when $\beta$ ranges over all rules  (\ref{eq:beta}). 
% \end{lemma}

\begin{proposition}\label{pr:cfl-decomposition}{(\cite{dlt2014}, Prop. 20)}
The language $L$ is a finite
union of languages of the form $C_{S}\diamond L''$ where % with the notation of (\ref{eqfla:rob}), 
$$
L''= 
w_{1}L_{A_{1}}w_{2}L_{A_{2}} \cdots w_{\ell}L_{A_{\ell}} w_{\ell+1},
$$
when $\beta$ ranges over all rules  (\ref{eq:beta}), and with respect to the notation (\ref{eqfla:rob}).
%
%With the previous notation $L$ is a finite
%union of languages of the form $C_{S}\diamond L''$ where
%$$
%L''= 
%w_{1}L_{A_{1}}w_{2}L_{A_{2}} \cdots w_{\ell}L_{A_{\ell}} w_{\ell+1}
%$$
\end{proposition}

%\begin{proof}
%In order to prove the inclusion of the right- 
%into left- hand side,  it suffices
%to consider $w=  \alpha u \beta,$ with $u\in L'(G) $ and $( \alpha, \beta)\in \G_S.$
%One has $S  \displaystyle\mathop{\Rightarrow}^{*}u$ and 
% $S  \displaystyle\mathop{\Rightarrow}^{*} \alpha S \beta$
%so that $S  \displaystyle\mathop{\Rightarrow}^{*} \alpha S \beta \mathop{\Rightarrow}^{*}  \alpha u \beta,$  and thus $w\in L$. 
%
%Let us prove the opposite inclusion. A word  $w\in L$ admits a derivation
% $S   \displaystyle\mathop{\Rightarrow}^{*} w $. 
% If the symbol $S$ does not occur in the derivation except 
% at the start of the derivation,  then   $w\in L'(G)$. Otherwise
% factor this derivation into  
% $S   \displaystyle\mathop{\Rightarrow}^{*} \alpha S \beta  \mathop{\Rightarrow}^{*} w $ 
%such that $S$ does not occur in the second part of the derivation
%except in the sentential form $\alpha S \beta$.
%%
%Reorder the derivation $\displaystyle\alpha S \beta  \mathop{\Rightarrow}^{*} w $
%into $\displaystyle\alpha S \beta \mathop{\Rightarrow}^{*} \gamma S \delta  \mathop{\Rightarrow}^{*} w $
%so that $\gamma,  \delta\in \Sigma^*$.
%This implies  $w=\gamma u \delta$ for some   $u\in L'(G)$,  
% which, by Lemma \ref{le:L'}, completes the proof.	\qed
%\end{proof}
    
           \subsection{Structuring of $\FIN $}\label{Sec:Structuring of FIN}
  We now recall a characterization, by
            Ginsburg and Spanier \cite{GinSpan} (see also Nivat \cite{Niva}), 
 of languages of $\FIN$, that will be used in Section \ref{subsec: the case of monoidal languages}.  
 To this purpose, we first recall that, given alphabets $\Sigma_1$ and $\Sigma_2$, and a family $\cal F$ of languages of $\Sigma_2^*$,
  a {\em $\cal F$-substitution} is a morphism $\theta : \Sigma_1^* \rightarrow \Pi(\Sigma_2^*)$ from the free monoid $\Sigma_1^*$ into the
  multiplicative monoid $\Pi(\Sigma_2^*)$ of subsets of the free monoid $\Sigma_2^*$ such that
  $$\forall \ x\in \Sigma_1 \quad  \theta (x) \in \cal F.$$ 
   Let us define, recursively, the family $\{Qrt(k)\}_{k\in \N}\ $ of {\em quasi-rational languages of rank $k$} 
  (or {\em bounded derivation languages of rank $k$}) as:
  \begin{equation}\label{eq:bounded-lan-GinSpan}
Qrt(k) = \left \{
\begin{array}{ll}
LIN  & \quad \mbox{if} \ k=1,     \\
 LIN \circ Qrt(k-1) &    \quad  \mbox{if} \ k> 1,
  \end{array}
\right.
\end{equation}
where $LIN \circ Qrt(k-1)$ denotes the family of all the languages obtained as the images of linear languages, {\em via} 
$Qrt(k-1)$-substitutions, $\ k> 1$, i.e.
 $$
   \{\theta (L) : L \in LIN,\, \quad  \forall \; x\in \Sigma_1 \;\;\; \theta(x) \in  Qrt(k-1)\}.
 $$

% The family $Qrt(k)$  is equivalently formulated in terms of  of {\em composition of grammars} (see \cite{Ber79}, Sec.~II.2). The following holds.
 The family $Qrt(k)$  is equivalently formulated in terms  of  the well known construct of {\em composition of grammars} 
 (such construct is recalled in detail in the Appendix). The following holds.
 \begin{theorem}{(\cite{GinSpan}, Theorem 4.2, cf \cite{Ber79}, Sec. VII.5, Theorem 5.2)}
 \label{th:fundamental-GinSpan} 
 Let $L$ be a context-free language and $k\geq 1$. Then $L\in Ind(k)$ if and only if   $L\in Qrt (k)$. Moreover, $L\in Qrt (k)$ if and only if there exists a grammar $G$ with $L=L(G)$ such that
$G$ is  a {\em composition} 
 $ {\cal G}_1 \circ   \cdots \circ {\cal G}_k$
of $k$ families of linear grammars.
    \end{theorem}
%
%  XXXXXXXXX
%
%
%
%
%
%
%The following holds  (\cite{GinSpan}, Theorem 4.2, cf also \cite{Ber79}, Sec. VII.5, Theorem 5.2). 
%\begin{theorem}{}
% \label{th:fundamental-GinSpan} 
% For every $k\geq 1$, $Ind(k) = Qrt (k).$
% \end{theorem}
%
%A {\em composition of $k$ families of  grammars} $({\cal G}_1,   \ldots,  {\cal G}_k),$ with $k\geq 1$ % and $\card({\cal G}_1)=1$, 
%will be  the grammar
% $$G = {\cal G}_1 \circ {\cal G}_2 \circ \cdots \circ {\cal G}_k,$$
% obtained by iterating $(k-1)$ times the operation (\ref{eq:bounded-lan-GinSpan-comp-rk2}) of composition of grammars  on the families
% of context-free grammars  ${\cal G}_1,   \ldots,  {\cal G}_k$ (if the operation is well defined).
%%   As a straightforward consequence of this, 
%   By (\ref{eq:bounded-lan-GinSpan}),   Theorem  \ref{th:fundamental-GinSpan} can be formulated as follows. 
% \begin{theorem} \label{th:fundamental-GinSpan-1} 
%Let $L$ be an arbitrary language of $Qrt(k), \ k\geq 1$. Then $L$ is generated by a {\em composition} of $k$ families of linear grammars.
%  \end{theorem}

  \begin{example}\label{main-ex}
Let $\cal L$ be the language over $ \Sigma_2 = \{a,b\}$ given by $${\cal L} = \{a^nb^n : n\in \N\}^*.$$ 
 Observe that $\cal L$ is the image of the language $\{x^n : n\in \N\}$, $\Sigma_1=\{x\}$, under the substitution
$\theta : \Sigma_1^* \rightarrow \Pi(\Sigma_2^*)$, with  $\theta (x) =  \{a^nb^n : n\in \N\}$. Hence $\cal L$ 
 is a context-free language of index $2$. Then $\cal L$ is generated by the composition
  $G = {\cal G}_1 \circ {\cal G}_2$ of the linear grammars ${\cal G}_1=\langle V_1, \Sigma_1, P_1, S \rangle$ and ${\cal G}_2 =\langle V_2, \Sigma_2, P_2, \sigma \rangle$ where 
 $$V_1=\{S\},\;  \Sigma_1=\{\sigma\},\; P_1 = \{S \rightarrow \varepsilon, \;  S \rightarrow \sigma S\},$$ 
 and 
     $$V_2=\{\sigma\}, \; \Sigma_2=\{a, b\}, \; P_2 = \{\sigma \rightarrow \varepsilon, \; \sigma \rightarrow a\sigma b\}.$$
   \end{example}

 \section{Context-free  languages and semialgebraic sets}
\label{subsec:cf-lin}
%
%
%Here we show that $\closure(\varphi(L))$ is effectively eventually
%definable for languages generated by linear grammars
%and rational quantum automata.
%
%We adopt the notation from Section \ref{sec:context-free} for   context-free grammars.
%% 
%We recall the following notion that will be used in the proof of the next result (see \cite{Ber79}). A subset of a monoid $M$ is
%\emph{regular} if it is recognized by some finite $M$-automaton
%which differs from an ordinary finite 
%nondeterministic automaton
%over the free monoid by the fact the 
%transitions are labeled by elements in $M$. A \scriptstyle T
 Now we find convenient to recall (fixing some details of the proof) a result (cf \cite{dlt2014}, Prop. 7), obtained from the decomposition of
Section \ref{Sec:Structuring of a context-free language}.  Such a result is relevant in this study since it states, for an arbitrary 
 context-free language $L$, that the closure $\Closure{\varphi(L)}$ of the image of  $L$, under the morphism (\ref{eq:main-morphism})  associated with a 
quantum automaton,  is semialgebraic. To this purpose, the following construction is crucial.
With an arbitrary nonterminal $A$ of $G$, associate the set $M_A$  defined as
 \begin{equation}\label{eq:crucial-monoid}
 M_A=\{\varphi(u_1)\oplus \varphi(u_2)^{  \scriptstyle T}\mid (u_1,u_2)\in C_{A}\},
%M_A=\{\varphi(u_1)\oplus \varphi(u_2)^{  \mbox {T}}\mid (u_1,u_2)\in C_{A}\}, 
\end{equation}
where $ \varphi(u_2)^{\scriptstyle T}$ denotes the  transpose  of  $\varphi (u_2)$. 
$M_A$ will be called the {\em monoid of cycles of $A$} since the following holds.
 
\begin{lemma}\label{crucial:monoid} 
Let $A$ be a nonterminal symbol of $G$. 
$M_A$ is a monoid and  its closure  $\Closure{M_A}$  is an algebraic group.   \end{lemma}
 
 \begin{proof}
Observe first that $M_A$ is a subsemigroup of the group $O_{n}\oplus O_{n}.$ % \simeq O_{2n}$. 
Indeed, 
if $\varphi(u_1)\oplus \varphi (u_2)^{\scriptstyle T}$ and $\varphi(v_1)\oplus \varphi (v_2)^{\scriptstyle T}$ are 
in $M_A$ then we have
\begin{equation}\label{eq0:crucial:monoid}
  (\varphi(u_1)\oplus \varphi (u_2)^{\scriptstyle T}) (\varphi(v_1)\oplus \varphi (v_2)^{\scriptstyle T})
= \varphi(u_1)\varphi(v_1)\oplus \varphi(u_2)^{\scriptstyle T}\varphi(v_2)^{\scriptstyle T},
\end{equation}
 where $(u_1,u_2), (v_1,v_2)\in C_A$. On the other hand,  $(u_1v_1, v_2u_2) \in C_A$ and the corresponding matrix of $M_A$ 
\begin{equation}\label{matr-2}
  \varphi(u_1v_1)\oplus \varphi(v_2u_2) ^{\scriptstyle T} 
   \end{equation}
   equals (\ref{eq0:crucial:monoid}) for  $ \varphi(v_2u_2) ^{\scriptstyle T} =  \varphi(u_2)^{\scriptstyle T}\varphi(v_2)^{\scriptstyle T}.$
Since $(\varepsilon, \varepsilon)\in C_A$, then $M_A$ is a monoid. 
The second statement comes from   Theorems \ref{th:semigroup-in-compact} and   \ref{th:topological-closure-of-monoid}.
%$(u_1v_1, v_2u_2) \in C_S$ and the corresponding matrix of $M_S$ is
%\begin{equation}\label{matr-2}
%  (\varphi(u_1v_1)\oplus \varphi^{\scriptstyle T}(v_2u_2)) =\varphi(u_1v_1)\oplus \varphi^{\scriptstyle T}(u_2)\varphi^{\scriptstyle T}(v_2).
%   \end{equation}
%The claim follows because the matrices of (\ref{eq0:crucial:monoid}) and (\ref{matr-2}) are equal.
%Since $(\varepsilon, \varepsilon)\in M_S$, then $M_S$ is a monoid. 
%The second statement comes from   Theorems \ref{th:semigroup-in-compact} and   \ref{th:topological-closure-of-monoid}.
%Finally by
% Theorems \ref{th:semigroup-in-compact} and   \ref{th:topological-closure-of-monoid},  
% $\Closure{M_S}$ is algebraic.
%Observe first that $M_S$ is a semigroup. Indeed, 
%if $\varphi(u_1)\oplus \varphi^{\scriptstyle T}(u_2)$ and $\varphi(v_1)\oplus \varphi^{\scriptstyle T}(v_2)$ are 
%in $M_S$ then we have
%%
%$$
%\varphi^{\scriptstyle T}(b) \varphi^{\scriptstyle T}(d)=\varphi(b)^{\scriptstyle T} \varphi(d)^{\scriptstyle T}
%=(\varphi(d) \varphi(b))^{\scriptstyle T}= \varphi(d b)^{\scriptstyle T}= \varphi^{\scriptstyle T}(d b)
%$$
%%
%which yields 
%%
%$$
%\begin{array}{l}
%(\varphi(u_1)\oplus \varphi^{\scriptstyle T}(u_2)) (\varphi(v_1)\oplus \varphi^{\scriptstyle T}(v_2))
%=\varphi(a c)\oplus \varphi^{\scriptstyle T}(db).
%\end{array}
%$$
%%
%Since $(\varepsilon, \varepsilon)\in M_S$, then $M_S$ is a monoid. 
%The second statement comes from   Theorems \ref{th:semigroup-in-compact} and   \ref{th:topological-closure-of-monoid}.
%%Finally by
%% Theorems \ref{th:semigroup-in-compact} and   \ref{th:topological-closure-of-monoid},  
\qed
  \end{proof}
\begin{proposition}\label{main-prop-dlt2014}  
 Let $L$ be a context-free language. The set $\Closure{\varphi(L)}$ is semialgebraic. 
Moreover, if for every $A\in V$,  $\Closure{M_A}$ is effective algebraic, then $\Closure{\varphi(L)}$ is effective semialgebraic. 
% Let $L$ be a context-free language. The closure $\Closure{\varphi(L)}$ of the image, under the morphism (\ref{eq:main-morphism}) 
%of  $L$ is semialgebraic. 
%Moreover, if for every $A\in V$,  $\Closure{M_A}$ is effective algebraic, then $\Closure{\varphi(L)}$ is effective semialgebraic. 
%
%
%Let $\varphi$ be the morphism of Eq. (\ref{eq:main-morphism}). If $L$ is generated by a context-free grammar, then 
%$\Closure{\varphi(L)}$ is algebraic. 
%Furthermore,  
%if the grammar is linear and  if the quantum automaton is rational
%then $\Closure{\varphi(L)}$  is effectively 
%eventually  definable and the $(L, Q)$ intersection
%is decidable.
\end{proposition}
 
\begin{proof}
By Proposition \ref{pr:cfl-decomposition}, the language $L$ is a finite
union of languages of the form
$C_{S}\diamond L'' $ with
\begin{equation}
\label{eqfla:claim2-a} 
L''= 
w_{1}L_{A_{1}}w_{2}L_{A_{2}} \cdots w_{\ell}L_{A_{\ell}} w_{\ell+1}
\end{equation}
where, for every  $1\leq i \leq \ell +1$, 
 $w_{i}\in \Sigma^*$ and $A_{i}\in V'$.
It suffices to show by induction on the number
of nonterminal symbols that the subsets  
%\bigskip
%
%\noindent
%{\bf        *** In formula (18) should be replaced by $ \Closure{(M_{S}\diamond \varphi(L'')}$ }\bigskip
%
%\noindent
%
%
\begin{equation}
\label{eq:L-double-prime}
 \Closure{\varphi(C_{S})\diamond \varphi(L'')} 
\end{equation}
are semialgebraic. %  in all cases and  effectively eventually definable when the quantum automaton is rational and the grammar of the language is linear.
If $\card(V) =1$, i.e.,  the set of nonterminal symbols is reduced to $S$ 
then $L$ is reduced to $C_S\diamond L'(G)$ and $L'(G)$ is finite. 
We may further assume that there is a sole
terminal rule $S\rightarrow w$, so that $L'(G)=\{w\}$.
By Theorem \ref{th:fundamental} one has\footnote{By Proposition \ref{pr:union-prod-phi}, 
if $\Closure{M_S}$ is algebraic, then $\{X\oplus Y^{\scriptstyle T} : X\oplus Y\in \Closure{M_S}\}$ is so} 
\begin{equation}
\label{eqfla:claim2-C} 
 \Closure{\varphi(L)}
=\{X\varphi(w)Y^{\scriptstyle T}  \mid X\oplus Y\oplus \{\varphi(w)\}\in \Closure{M_S\oplus \varphi(w)}\}.
\end{equation}
%M_S
%By Corollary \ref{cor:closure-of-product} we have
%
On the other hand, one has % so that % we have
$$\Closure{M_S\oplus \varphi(w)}
= \Closure{M_S}\oplus \Closure{\varphi(w)}
= \Closure{M_S}\oplus \varphi(w)
$$
which, by Lemma \ref{crucial:monoid} and Proposition \ref{pr:blocks-of-product},
is algebraic. 
By (\ref{eqfla:claim2-C}) and Proposition~\ref{pr:product-of-blocks}, we conclude that  
$ \Closure{\varphi(L)}$ is effective semialgebraic. The basis of the induction is thus proved. 
% , resp. effectively eventually definable.
%In that latter case the $(L, Q)$ intersection is decidable.
Now assume $V$ contains more than one nonterminal symbol.
%We first prove that for each  nonterminal symbol $A$, 
%%
%$
%\Closure{\varphi(C_{S}\diamond L_{A})}
%$ 
%is  algebraic.
% in the general case 
%and effectively eventually definable when the grammar is 
%linear and the quantum automaton is rational.
By Theorem \ref{th:fundamental}, Corollary \ref{cor:closure-of-product} and Eq. (\ref{eqfla:claim2-a}), we get
% $\Closure{\varphi(C_{S}\diamond L'')}$
%is the subset 
%
%
%
%
%
%\bigskip
%
%\noindent
%{\bf        *** The same consideration in formula (18) holds HERE }\bigskip
%
%\noindent
%
%
%
%
$$
%\Closure{\varphi(C_{S}\diamond L'')}  
\Closure{\varphi(C_{S})\diamond \varphi(L'')} 
=\{XZY^{\scriptstyle T}\mid X\oplus Y\oplus Z\in \Closure{M_S}\oplus \Closure{\varphi(L'')}\} =
$$
% %
% with $L''$ as in  (\ref{eqfla:claim2-a}),
% i.e.,
% %
$$
\{XZY^{\scriptstyle T}\mid X\oplus Y\oplus Z \in \Closure{M_S}\oplus \Closure{\varphi(w_{1})\varphi(L_{A_{1}})\cdots 
\varphi(w_{\ell})\varphi(L_{A_{\ell}})\varphi(w_{\ell+1})}\}
$$
 By Corollary  \ref{cor:closure-of-product},
we have

$$
\begin{array}{ll}
& \Closure{\varphi(w_{1})\varphi(L_{A_{1}})\cdots 
\varphi(w_{\ell})\varphi(L_{A_{\ell}})\varphi(w_{\ell+1}))\\
= &\varphi(w_{1})\Closure{\varphi(L_{A_{1}})}\cdots 
\varphi(w_{\ell})\Closure{\varphi(L_{A_{\ell}})}\varphi(w_{\ell+1})}
\end{array}
$$
which shows, via Proposition \ref{pr:union-prod-phi}
and by the induction hypothesis,
that this subset is semialgebraic. 
% , resp. effectively, eventually definable. 
Then  its direct sum with
$\Closure{M_S}$ is semialgebraic.
% and
%effectively,
%eventually definable if the grammar is
%linear and the quantum automaton is rational. 
We conclude by applying Proposition \ref{pr:product-of-blocks}. 
%
%We first prove that for each  nonterminal symbol $A$, 
%%
%$
%\Closure{\varphi(C_{S}\diamond L_{A})}
%$ 
%is  algebraic.
%% in the general case 
%%and effectively eventually definable when the grammar is 
%%linear and the quantum automaton is rational.
%By Theorem \ref{th:fundamental} 
%and  Corollary \ref{cor:closure-of-product},
% $\Closure{\varphi(C_{S}\diamond L'')}$
%is the subset 
%%
%$$
%\Closure{\varphi(C_{S}\diamond L'')}
%=\{XZY^{\scriptstyle T}\mid X\oplus Y\oplus Z\in \Closure{M_S}\oplus \Closure{\varphi(L'')}\} =
%$$
%% %
%% with $L''$ as in  (\ref{eqfla:claim2-a}),
%% i.e.,
%% %
%$$
%\{XZY^{\scriptstyle T}\mid X\oplus Y\oplus Z \in \Closure{M_S}\oplus \Closure{\varphi(w_{1})\varphi(L_{A_{1}})\cdots 
%\varphi(w_{\ell})\varphi(L_{A_{\ell}})\varphi(w_{\ell+1})}\}
%$$
% %
% By Cororally  \ref{cor:closure-of-product}
%we have
%%
%$$
%\begin{array}{ll}
%& \Closure{\varphi(w_{1})\varphi(L_{A_{1}})\cdots 
%\varphi(w_{\ell})\varphi(L_{A_{\ell}})\varphi(w_{\ell+1}))\\
%= &\varphi(w_{1})\Closure{\varphi(L_{A_{1}})}\cdots 
%\varphi(w_{\ell})\Closure{\varphi(L_{A_{\ell}})}\varphi(w_{\ell+1})}
%\end{array}
%$$
%which shows, via Proposition \ref{pr:union-prod-phi}
%and by induction hypothesis 
%that this subset is algebraic. 
%% , resp. effectively, eventually definable. 
%Then  its direct sum with
%$\Closure{M_S}$ is algebraic.
%% and
%%effectively,
%%eventually definable if the grammar is
%%linear and the quantum automaton is rational. 
%We conclude by applying Proposition \ref{pr:product-of-blocks}. 
The second part of the statement is obtained from the hypothesis by following  the same proof. 
\qed  \end{proof}
  
 \begin{lemma}\label{main-lm-lin:dlt2014}
Assume that  $L$ is a context-free linear language and the quantum automaton is rational. Then, for every variable $A$, $\Closure{M_A}$ is effective algebraic.
%Let $A$ be a nonterminal symbol of $G$.  If $L$ is a context-free linear language and the quantum automaton is rational, then $\Closure{M_A}$ is effective algebraic.
 \end{lemma}
 
 \begin{proof}
 Following  \cite{Ber79},  let us first recall that a  subset of a monoid $M$ is %  said to be 
\emph{regular} if it is recognized by some finite $M$-automaton,
which differs from an ordinary finite 
nondeterministic automaton
over the free monoid by the fact the 
transitions are labeled by elements in $M$. 
Further, by \cite{Anisimov}, the subgroup generated 
by a regular subset of a monoid has an 
effective finite generating set.
Observe now that,  if $L$ is a context-free linear language, then $M_A$  is a regular submonoid
of  the group of orthogonal matrices $O_{n}\oplus  O_{n}$. Indeed, it is recognized by the finite 
$O_{2n}$-automaton whose states
are the nonterminal symbols, the transitions
are of the form $B\trans{\varphi(a)\oplus  \varphi(b) ^{\scriptstyle T}} C,$ where 
$B\rightarrow aCb$ is a rule of the grammar
and where the initial and final states coincide 
with $A$.   Hence $\langle M_A \rangle$  has an 
effective finite generating set and the claim follows
by applying to $\langle M_A \rangle$ the algorithm of  Derksen et al. \cite{derksen}. 
%By following  \cite{Ber79},  let us first recall that a  subset of a monoid $M$ is said to be 
%\emph{regular} if it is recognized by some finite $M$-automaton,
%which differs from an ordinary finite 
%nondeterministic automaton
%over the free monoid by the fact the 
%transitions are labeled by elements in $M$. 
%According to the last definition,  if $L$ is a context-free linear language, then $M_A$  is a regular submonoid
%of  the group of orthogonal matrices $O_{n}\oplus  O_{n}$. Indeed, it is recognized by the finite 
%$O_{2n}$-automaton whose states
%are the nonterminal symbols, the transitions
%are of the form $A\trans{\varphi(a)\oplus  \varphi(b) ^{\scriptstyle T}} B,$ where 
%$A\rightarrow aBb$ is a rule of the grammar
%and where the initial and final states coincide 
%with $S$.  Now, by \cite{Anisimov}, the subgroup generated 
%by a regular subset of a monoid has an 
%effective finite generating set, which implies that $\langle M_A \rangle$  has this property.    The claim finally follows
%by applying to $\langle M_A \rangle$ the algorithm of  Derksen et al. \cite{derksen}. 
\qed  \end{proof}
 Proposition \ref{main-prop-dlt2014}   together with Lemma \ref{main-lm-lin:dlt2014} finally yield.
  \begin{corollary}{(\cite{dlt2014}, Prop. 22)}
\label{main:dlt2014}
If $L$ is a context-free linear language and the quantum automaton $\cal Q$  is rational, then its closure  $\Closure{\varphi(L)}$  is an effective semialgebraic set.
As a consequence, it is recursively decidable  whether or not  
$L\cap |{\cal Q}_{>}|=\emptyset$ holds.
  \end{corollary}
An intermediate class of languages in between the linear and the finite index ones, is that of {\em metalinear} (or {\em ultralinear}) languages $\META$.  By definition, a language $L$ is in $\META$ if  
  there exist a context-free grammar $G$ with $L=L(G)$  and an integer $k\in \N$ such that, in every  derivation  of $G$,  the number of occurrences of variables in each sentential form  
   does not exceed $k$. The following corollary holds.
 
   \begin{corollary}
\label{main:dlt2014-bis}
If $L\in \META$ and  $\cal Q$ is a  rational quantum automaton, then
it is recursively decidable  whether or not  
$L\cap |{\cal Q}_{>}|=\emptyset$ holds, 
%It is recursively decidable  whether or not  
%$L\cap |{\cal Q}_{>}|=\emptyset$ holds, if $L\in \META$ and  $\cal Q$ is a  rational quantum automaton.
  \end{corollary}

 \begin{proof}
 It is known (see \cite{Ber79,Gins}) that  $L$ is in $\META$ if and only if  $L$ is a finite union of finite products of context-free linear languages.
 Thus the claim follows from  Corollary \ref{main:dlt2014}, the closure properties of Section \ref{Sec:Closure properties} and Proposition \ref{pr:meta-resuositionlt}.
  \qed  \end{proof}

 \section{Main results}\label{sec: main results}
We will prove now that the problem is decidable for two
families of languages: 
the languages generated by the class of  restricted  matrix context-free grammars and, subsequently, the languages generated by context-free  monoidal grammars.

\subsection{The case of finite index matrix  languages} \label{extens-BSL}
 Now we propose a first non trivial extension   of the main results of decidability provided in \cite{dlt2014}. 
  Such an extension is based upon the machinery of 
 Section~\ref{subsec:cf-lin}, and some combinatorial tools developed by Choffrut, Varricchio and the second author in  \cite{tcsCDV2010} for the study of the
 counting functions of 
 rational relations.
 To this purpose, following \cite{dp-matr-gr}, Ch. $1$, Sec. $1.1$, we first recall that a {\em matrix context-free  grammar} 
 (which we will henceforth simply call  {\em matrix grammar}) 
 is a tuple $G=\langle V, \Sigma, M, S \rangle$, where $V$, $\Sigma$, and $M$ are respectively the finite sets of nonterminals, terminals, and matrices (of rules), and $S \in V$ is the start symbol.
 Each {\em matrix}  $m \in M$ is a finite sequence $m = (p_1, \ldots, p_s)$ where each $p_i$, $1 \leq i \leq s$, is
a context-free production from $V$ to $(V\cup \Sigma)^*$. 
We denote by $P_M$ the set of all the context-free productions that appear in the matrices of $M$.
For an arbitrary $m\in M$, we define the {\em $1$-step derivation in $G$,} $x \derivestep[m] y,$ with $x,y \in (V\cup \Sigma)^*$ if $$x= x_0 \derivestep[p_1]x_1
\derivestep[p_2] \cdots \derivestep[p_s] x_s = y,$$ where $\derivestep[p_i]$ denotes the standard context-free derivation relation, defined by the rule $p_i$.
This is equivalent to say that  $x= x_0, x_s = y$, and, for every $i=1, \ldots, s,$
$$x_{i-1}= x'_{i-1}X_{i}x''_{i-1}, \quad x_{i}= x'_{i-1}\alpha_{i}x''_{i-1},$$ where  $p_i=(X_{i}\rightarrow \alpha_{i})$, and 
for some $x'_{i-1}, x''_{i-1}\in (V\cup \Sigma)^*$.
In other words, a $1$-step derivation in the matrix grammar $G$ corresponds to a $s$-step derivation in the context-free grammar $\langle V, \Sigma, P_M, S \rangle$,
i.e., using, one by one, all  the context-free productions of $m$ (w.r.t. the order the $p_i$'s appear in $m$).
 Let $M^*$ be the set of the finite sequences of matrices and $\alpha \in M^*$.
If $\alpha = \varepsilon$, then $x \Rightarrow_{\alpha} x, x \in V^*$; if one has
\[
\alpha=m_1m_2\cdots 	m_s
\mbox{\quad and\quad}
x_0\derivestep[m_1]x_1\derivestep[m_2]\cdots \ x_{s-1}\derivestep[m_s]x_s,
\]
with $x_i\in (V\cup \Sigma)^*,$ $m_j\in M$, $1\leq j\leq s$, $0\leq i\leq s$, then we write $x_0\derivestep[\alpha]x_s$.
The  \emph{language generated} by $G$ is $L(G) = \{w \in  \Sigma^* \mid S \derivestep[\alpha] w\}$.
 
The following definition is instrumental. It corresponds to a specific case of the {\em simple matrix grammars} introduced by Ibarra in \cite{ibarra-sml} (see also \cite{ibarra-mcquillan}).
  \begin{definition}
 \label{defo-gram-matr}
  A matrix grammar $G=(V,\Sigma,M,S)$ is said to be {\em restricted of index $k$}, or simply {\em restricted}, if the set of nonterminals of $G$ has the form
$V = \{S\} \cup V'$, where $S\notin V'$ and $V' = V_1 \cup \dots \cup V_k,$ where $ V_i \cap V_j = \emptyset,$ for all $i\neq j$, and a matrix of rules of $G$ can be only of one 
of the following forms
% A matrix grammar $G=(V,\Sigma,M,S)$ is said to be {\em restricted linear of index $k$}, or simply restricted, if the set of nonterminals of $G$ has the form
%$V = \{S\} \cup V'$, where $V' = V_1 \cup \dots \cup V_k,$ and $ V_i \cap V_j = \emptyset,$ for all $i\neq j$, and a matrix rule of $G$ can be only of one 
%of the following forms

 \begin{itemize}
\item $(X_1 \rightarrow u_1Y_1v_1, \ldots, X_k \rightarrow u_kY_k v_k)$, for some $X_i, Y_i \in V_i, \ u_i, \ v_i \in \Sigma^*$, with $1\leq i\leq k$,

\item $(X_1 \rightarrow \varepsilon, \ldots, X_k \rightarrow \varepsilon)$, with $X_i\in V_i,$ $1\leq i\leq k$,
%\item For every $X_i\in V_i$ and $1\leq i\leq k$, $(X_1 \rightarrow \varepsilon, \ldots, X_k \rightarrow \varepsilon)$.

\item $(S\rightarrow X_1\cdots X_k)$, for some $X_i\in V_i$, with $1\leq i\leq k$.
\end{itemize}
 \end{definition}
 Such grammars are very akin to {\em simple matrix linear grammars}  (cf \cite{ibarra-sml}, pag. 363,  \cite{dp-matr-gr}, Ch. $1$, Sec. $5$). %Definition 1.5.1).
    \begin{example}\label{first-example-matrix-gramm}
   Let $L = \{a^nb^nc^n: n\in \N\}$ be the language over the alphabet $\Sigma = \{a,b, c\}$.
   Let $G$ be the restricted matrix grammar with the matrices of rules
   $m_1=(S\rightarrow ABC)$, $m_2=(A \rightarrow aA, B \rightarrow bB, C \rightarrow cC),$ and
 $m_3=(A \rightarrow~\varepsilon, B \rightarrow~\varepsilon, C \rightarrow \varepsilon).$
 Note that the other components of $G$ can be deduced from these matrices.
 One easily checks that $L=L(G)$. 
 Indeed, each word $a^nb^nc^n, n\geq0,$ can be generated by using first the matrix $m_1$, then the matrix~$m_2$ $n$ times and finally the matrix $m_3$.
 Hence $L\subseteq L(G)$. The reverse inclusion is proved similarly.
%       Let $L_k$, with $k\in \N$,  be the language over $\Sigma = \{a,b\}$ given by $L_k = \{u^k : u\in \Sigma^*\}$.
%  It is easily checked that $L_k$ can be generated by the restricted grammar $G=\langle V, \Sigma, M, S \rangle$, 
%  where $V = \{S, X_1, \ldots, X_k\}$, and $M$ is given by the matrix rules $(S\rightarrow X_1  \cdots X_k)$,  
% $(X_1 \rightarrow \varepsilon, \ldots, X_k\rightarrow \varepsilon)$, and, for every $\sigma \in \Sigma$,
% $(X_1 \rightarrow \sigma X_1, \ldots, X_k \rightarrow \sigma X_k)$.
        \end{example}

The main result of this section is the following.
  \begin{proposition}\label{main-prop-gen-matr-gramm-bsl}   
 %   If $L$ is a language  generated by a restricted grammar, then   one has: 
Let $L$ be a language generated by a restricted grammar and let $\cal Q$ be a quantum automaton. Then  one has:
  \begin{enumerate}
\item[i.]  $\Closure{\varphi(L)}$ is semialgebraic.
\item[ii.] If the quantum automaton ${\cal Q}$ is rational, then $\Closure{\varphi(L)}$ is effective semialgebraic and
the $(L, Q)$ Intersection problem
is decidable.
 \end{enumerate}

    \end{proposition}

     \begin{proof} Let $L$ be a language generated by a grammar as in Definition \ref{defo-gram-matr}. Let us prove
           (i). Set ${\cal X} = (V')^k$, and consider the sets 
           $${\cal Y} = \{X_1\cdots X_k\in {\cal X} : \exists \ m \in M : m=(S\rightarrow X_1\cdots X_k) \},$$ and 
           $${\cal Z}  = \{X_1\cdots X_k\in {\cal X} : \exists \ m \in M : m=(X_1 \rightarrow \varepsilon, \ldots, X_k \rightarrow \varepsilon)\}.$$
 
           Denote by $(\Sigma^*)^{2k}$ the multiplicative monoid of $2k$-fold relations on $\Sigma^*$.
           It is easily checked that, given $w\in \Sigma^*$,
       $w\in L$ if and only if there exists some 
       $\omega = (u_1,v_1, u_2,v_2, \ldots, u_k, v_k)\in (\Sigma^*)^{2k},$ such that $$w = u_1v_1u_2v_2 \; \cdots\; u_kv_k,$$ 
       and $\omega$ belongs to the set $\cal L$ defined as
   \begin{equation}
\label{eq:new-proof-bsl-2024-0}
{\cal L} = \bigcup_{p\in {\cal Y}, \; q\in \cal Z}\; {\cal L} _{pq},
\end{equation}
where, for $p = X_1\cdots X_k, \ q= Y_1\cdots Y_k$, ${\cal L} _{pq}$ is the subset of $(\Sigma^*)^{2k}$ defined as%  stands for 
$${\cal L} _{pq} =  \{\; ( u_1,v_1, \ldots, u_k, v_k)\;  : \;  \exists \ \alpha \in M ^* :\ X_1\cdots X_k \Rightarrow_{ \alpha}  u_1Y_1v_1\cdots u_kY_kv_k\; \}.$$

%      (i) Set ${\cal X} = (V')^k$ and  ${\cal Y} = \{X_1\cdots X_k\in {\cal X} : \exists \ m \in M : m=(S\rightarrow X_1\cdots X_k) \}$. It is easily checked that, given $w\in \Sigma^*$,
%       $w\in L$ if and only if there exists some 
%       $\omega = (w_1, \ldots, w_k)\in (\Sigma^*)^k,$ such that $w = w_1\cdots w_k$ 
%       and $\omega$ belongs to the set $\cal L$ defined as
%   \begin{equation}
%\label{eq:new-proof-bsl-2024-0}
%{\cal L} = \bigcup_{p\in {\cal Y}, \; q\in \cal X}\; {\cal L} _{pq},
%\end{equation}
%where, for $p = X_1\cdots X_k, \ q= Y_1\cdots Y_k$, ${\cal L} _{pq}$ is the subset of $(\Sigma^*)^k$ defined as%  stands for 
%$${\cal L} _{pq} =  \{(u_1, \ldots, u_k) : \ \exists \ \alpha \in M ^* :\ X_1\cdots X_k \Rightarrow_{ \alpha}  u_1Y_1\cdots u_kY_k\}.$$
 Let $s=(q_{1},\ldots, q_{n})$ be a
repetition-free
 sequence of elements of $\cal X$,
{i.e.}, for every positive integers $i, j$,
if $i\neq j$, then
$q_{i}\neq q_{j}$.
Clearly, $n$ is less than or equal to
the cardinal number of $\cal X$.
Denote by $\cal S$  the set
  of all such repetition-free sequences and observe that this set is finite.
Further, for every $p, q\in \cal X$, let ${\cal S}_{pq}$ be the set
$${\cal S}_{pq} =\{ s\in {\cal S}\;\mid\;
s=(q_{1},\ldots, q_{n}),\; q_{1}=p, \;q_{n}=q\}.
$$
Now, for every $p = X_1\cdots X_k, q= Y_1\cdots Y_k$, % for every $p, q\in \cal X$, 
define  the subset  $E_{pq}$ of $(\Sigma^*)^{2k}$ as:
$$E_{pq} =  \{\; ( u_1,v_1, \ldots, u_k, v_k)\;  : \; \exists \ m \in M  :\ X_1\cdots X_k \Rightarrow_{m}  u_1Y_1v_1\cdots u_kY_kv_k\; \},$$
%
%   \begin{equation}
%\label{eq:new-proof-bsl-2024-2}
%E_{pq} =  \{(u_1, \ldots, u_k) : \ \exists \ m \in M  :\ X_1\cdots X_k \Rightarrow_{m}  u_1Y_1\cdots u_kY_k\},
%\end{equation}
and the subset $C_{p}$ of $(\Sigma^*)^{2k}$   as:
%$$C_{q} =  \{(u_1, \ldots, u_k) : \ \exists \ \alpha \in M^*  :\ X_1\cdots X_k \Rightarrow_{\alpha}  u_1X_1\cdots u_kX_k\}.$$
%
$$C_{p} =   \{\; ( u_1,v_1, \ldots, u_k, v_k)\;  : \; \exists \ \alpha  \in M^*  :\ X_1\cdots X_k \Rightarrow_{\alpha}  u_1X_1v_1\cdots u_kX_kv_k\; \}.$$

%   \begin{equation}
%\label{eq:new-proof-bsl-2024-4}
%C_{q} =  \{(u_1, \ldots, u_k) : \ \exists \ \alpha \in M^*  :\ X_1\cdots X_k \Rightarrow_{\alpha}  u_1X_1\cdots u_kX_k\}.
%\end{equation}
%One easily checks that $C_{q}$ is a submonoid of the monoid $(\Sigma^*)^{2k}$.
%
For every $s=(q_{1},\ldots, q_{n})\in \cal S$,  define the (possibly empty) subset  ${\cal L}_{s}$ of $(\Sigma^*)^{2k}$   as:
\begin{equation}
\label{eq:new-proof-bsl-2024-3}
{\cal L}_{s} = C_{q_{1}} \diamond E_{q_{1}q_{2}} \diamond C_{q_{2}} \diamond \cdots
\diamond C_{q_{n-1}}\diamond E_{q_{n-1}q_{n}}\diamond C_{q_{n}},
 \end{equation}
%
%\begin{equation}
%\label{eq:new-proof-bsl-2024-3}
%{\cal L}_{s} = C_{q_{1}} E_{q_{1}q_{2}}C_{q_{2}}\cdots
%C_{q_{n-1}}E_{q_{n-1}q_{n}}C_{q_{n}}
% \end{equation}
%where, for arbitrary subsets  $ {\cal M},  {\cal M}'$ of  $(\Sigma^*)^{2k}$, the set  ${\cal M}  \diamond  {\cal M} ' $ is defined as
%  $$
% {\cal M}  \diamond  {\cal M} ' = \{\ (uu',\ v'v) \ \mid (u,v)\in {\cal M} \mbox{ and } (u',v')\in {\cal M}'\ \}.
%$$
%where, for arbitrary subsets  $ {\cal M},  {\cal M}'$ of  $(\Sigma^*)^{2k}$, the set  ${\cal M}  \diamond  {\cal M} ' $ is defined as
%  $$
% {\cal M}  \diamond  {\cal M} ' = \{\ (uu',\ v'v) \ \mid (u,v)\in {\cal M} \mbox{ and } (u',v')\in {\cal M}'\ \}.
%$$
%
%  \label{eq:sandwich}
where $(\diamond)$ corresponds to the extension of the operation (\ref{eq:sandwich-op}) to $(\Sigma^*)^{2k}$, that is,
 for subsets  $ {\cal M},  {\cal M}'$ of  $(\Sigma^*)^{2k}$, the tuple $( w_1, \ldots, w_{2k}) \in {\cal M}  \diamond  {\cal M} ' $ if and only if
 there exist $$( u_1,v_1, \ldots, u_k, v_k)\in {\cal M}\;\; \mbox{ and }\;\; ( u'_1,v'_1, \ldots, u'_k, v'_k)\in {\cal M}'$$ such that,
 for every $i=1, \ldots, k,$
  $$
 w_{2i - 1 } \ =  \ u _i u' _i \;\;  \;\; \mbox{ and }\;\; \;\;    w_ {2i} = v ' _i v_i.
$$
%   $$
% w_{1 } \ =  \ u _1 u' _1, \ w_ {2} = v ' _1 v_1, \ w_{3 } \ =  \ u _3 u' _3, \ w_ {4} = v ' _3 v_3,     w_{2k - 1 }  \ =  \ u _k u' _k, \ w_ {2k} = v ' _k v_k.
%$$

  \label{eq:sandwich}

 By using an argument similar to that of \cite{tcsCDV2010}, Theorem 2, one checks  that 
   \begin{equation}
\label{eq:new-proof-bsl-2024-4}
{\cal L}_{pq} = \bigcup_{s\in {\cal S}_{pq}} \ {\cal L}_{s}.
\end{equation}
Let now consider the monoid morphism 
  $$\widehat{\varphi} : (\Sigma^*)^{2k} \rightarrow O_{2nk},$$ 
 which maps every element $\omega= (u_1,v_1,  \ldots, u_k, v_k)$ into the orthogonal matrix 
  $$\widehat{\varphi}(\omega) = \varphi (u_1)\ \oplus\  \varphi (v_1) \ \oplus\  \cdots \ \oplus\   \varphi (u_k)\ \oplus\  \varphi (v_k).$$ %  of the $2k$-fold product $O_{2nk}$ of $O_{n}.$
% $$\widehat{\varphi}(\omega) = \varphi (u_1)\ \oplus\  \varphi (v_1) \ \oplus\  \varphi (u_2)\ \oplus\  \varphi (v_2)\ \oplus\  \cdots \ \oplus\   \varphi (u_k)\ \oplus\  \varphi (v_k),$$ of the $2k$-fold product of $O_{n}.$

Since, by Theorem \ref{th:fundamental}, one has
$$\Closure{{\varphi}(L)} \ = \ \{M_1\cdots M_{2k} \ : M_1 \oplus \cdots  \oplus M_{2k}\in \Closure{\widehat{\varphi}(\cal L)}  \},$$
to show that $\Closure{{\varphi}(L)}$ is semialgebraic,
by Proposition~\ref{pr:product-of-blocks}, it is enough to show that $\Closure{\widehat{\varphi}(\cal L)}$ is so. 
%Now we prove that $\Closure{{\varphi}(L)}$ is semialgebraic. 
%Since, by Theorem \ref{th:fundamental}, one has
%$$\Closure{{\varphi}(L)} \ = \ \{M_1\cdots M_{2k} \ : M_1 \oplus \cdots  \oplus M_{2k}\in \Closure{\widehat{\varphi}(\cal L)}  \},$$
%by Proposition~\ref{pr:product-of-blocks}, it is enough to show that $\Closure{\widehat{\varphi}(\cal L)}$ is so.
%
%
%
%
To this purpose, with an arbitrary $q\in {\cal X}$, associate the subset $M_q$ of $O_{2nk}$  defined as
% \begin{equation}\label{eq:crucial-matrix-monoid}
% M_q= \{\ \varphi (u_1)\  \oplus \ \varphi (v_1)^{\scriptstyle T} \  \oplus\   \cdots  \oplus\   \varphi (u_k)\ \oplus\  \varphi (v_k)^{\scriptstyle T}
%\end{equation}
%where $(u_1,v_1, \ldots, u_k, v_k)\in C_{q}$.
 \begin{equation}\label{eq:crucial-matrix-monoid}
 \{ \varphi (u_1) \oplus  \varphi (v_1)^{\scriptstyle T}  \oplus   \cdots  \oplus   \varphi (u_k)\ \oplus\  \varphi (v_k)^{\scriptstyle T}    \mid    (u_1,v_1, \ldots, u_k, v_k)\in C_{q}\}.
\end{equation}
%\[
%\left \{
%\begin{array}{lll}
%\varphi (u_1) \oplus  \varphi (v_1)^{\scriptstyle T}  \oplus   \cdots  \oplus   \varphi (u_k)\ \oplus\  \varphi (v_k)^{\scriptstyle T}  &  &   \\
% \quad \quad  \quad \quad  \mid    \quad \quad   \quad \quad   \quad \quad   (u_1,v_1, \ldots, u_k, v_k)\in C_{q}     & & 
%\end{array}
%\right \}
%\]

% \[
%\left \{
%\begin{array}{lll}
%\varphi (u_1)\  \oplus  \ \varphi (v_1)^{\scriptstyle T} \ \oplus \  \varphi (u_2)\ \oplus\  \varphi (v_2)^{\scriptstyle T}\ \oplus\   \cdots  \oplus\   \varphi (u_k)\ \oplus\  \varphi (v_k)^{\scriptstyle T}  &  &   \\
% \quad \quad  \quad \quad  \mbox{such that}    \quad \quad   \quad \quad   \quad \quad   (u_1,v_1, \ldots, u_k, v_k)\in C_{q}     & & 
%\end{array}
%\right \}
%\]

  By proceeding as in Lemma \ref{crucial:monoid}, one proves that 
 $M_q$ is a monoid and  its closure  $\Closure{M_q}$  is an algebraic group.  % $M_q$ will be still called the {\em (generalized) monoid of cycles of $q$}.
From this, by   Proposition \ref{pr:union-prod-phi}, it follows that $\Closure{\widehat{\varphi}({C}_q)}$ is an algebraic set. 
 This implies, via  Theorem \ref{th:fundamental} and Proposition~\ref{pr:sandwich},
 that  every set $\Closure{\widehat{\varphi}({\cal L}_s)}$ of (\ref{eq:new-proof-bsl-2024-3}) is semialgebraic.
 Then the semialgebraicity of $\Closure{\widehat{\varphi}(\cal L)}$ follows, by applying Proposition~\ref{pr:union-prod-phi}, from the last condition and  (\ref{eq:new-proof-bsl-2024-0}), (\ref{eq:new-proof-bsl-2024-4}).
  
\medskip

\noindent
(ii) Assume that  the quantum automaton ${\cal Q}$ is rational. 
First we prove that  $\Closure{M_q}$ is effective algebraic.
One proceeds as in the proof of
 Lemma \ref{main-lm-lin:dlt2014}. 
% Indeed, observe first that, since $C_q$ is a submonoid of $(\Sigma^*)^k$
% and ${\widehat{\varphi}}$ is a morphism, then ${\widehat{\varphi}}({C}_q)$ is  a submonoid of $O_{nk}$ as well.
 Indeed,  since $L$ is generated by a restricted matrix grammar, then $M_q$ is  a regular submonoid of $O_{2nk}$.
Precisely, it is recognized by the finite 
$O_{2nk}$--automaton whose states
are the elements of $\cal X$, the initial and final states coincide 
with $q$, and the transitions
are of the form $$e=(s\trans{\ell(e)} t),\quad s, t \in \cal X,$$ 
where, being $s= X_1\cdots X_k, \ t = Y_1\cdots Y_k$, there exists $m\in M$ such that
$$X_1\cdots X_k \Rightarrow_{m}   u_1Y_1v_1\cdots u_kY_kv_k,$$
with $\ell (e)\in O_{2nk}$ defined as
$$\ell (e) =  \varphi (u_1) \oplus  \varphi (v_1)^{\scriptstyle T}  \oplus   \cdots  \oplus   \varphi (u_k)\ \oplus\  \varphi (v_k)^{\scriptstyle T} .$$
  Hence $M_q$  is a regular monoid and the group $\langle M_q \rangle$ has an 
effective finite generating set. 
By applying to $\langle M_q \rangle$ the algorithm of  Derksen et al. %  \cite{derksen}, 
one gets an effective presentation of $\Closure{M_q}$. 
From this, by   Proposition \ref{pr:union-prod-phi}, it follows that $\Closure{\widehat{\varphi}({C}_q)}$ is effective algebraic as well. 
Finally, since all the steps of the proof (i) are effective, the fact that $\Closure{\widehat{\varphi}({C}_q)}$ is  effective algebraic
implies that $\Closure{{\varphi}(L)}$ is effective semialgebraic as well. 
The claim follows by applying Proposition~\ref{pr:meta-resuositionlt}.
%  Hence ${\widehat{\varphi}}({C}_q)$  is regular and $\langle {\widehat{\varphi}}({C}_q) \rangle$ has an 
%effective finite generating set. 
%By applying to $\langle {\widehat{\varphi}}({C}_q) \rangle$ the algorithm of  Derksen et al. \cite{derksen}, one gets an effective presentation of $\Closure{\widehat{\varphi}({C}_q)}$. 
%Finally, since all the steps of the proof (i) are effective, the fact that $\Closure{\widehat{\varphi}({C}_q)}$ is  effective algebraic
%implies that $\Closure{{\varphi}(L)}$ is effective semialgebraic as well. 
%The claim follows by applying Proposition~\ref{pr:meta-resuositionlt}.
 \qed  \end{proof}
 \noindent
 We now discuss some straightforward corollaries of  Proposition \ref{main-prop-gen-matr-gramm-bsl}.  
 First, remark that, by applying the proposition to the matrix grammars of index $k=1$, we recover \cite{dlt2014}, Prop. 22 (cf Corollary \ref{main:dlt2014}).
 Moreover, note that the language of Example \ref{first-example-matrix-gramm} is an instance of the following more general result. 
 To this purpose, we recall that a  \emph{bounded semilinear}  language
is a finite union of \emph{linear} languages which are languages of  the form
\begin{equation}
\label{eq:linear}
L=\{u_{1}^{n_{1}}\cdots u_{k}^{n_{k}} \mid(n_{1},\ldots, n_{k}) \in {R}\},
\end{equation}
for some fixed words $u_{i}\in \Sigma^*$ for $i=1, \ldots, k$ and $R\subseteq \N^k$
is a linear set, i.e., there exists ${\bf v}_{0}, {\bf v}_{1}, \ldots, {\bf v}_{\ell}\in \N^k$
such that 
\begin{equation}
\label{eq:linear-vectors}
R=\{{\bf v}_{0}+ \lambda_{1} {\bf v}_{1} + \cdots + \lambda_{\ell} {\bf v}_{\ell}  \mid \lambda_{1},  \ldots,  \lambda_{\ell}\in \N\}.
\end{equation}

  \begin{lemma}\label{gen-matr-gramm-bsl} 
 Every bounded semi-linear language can be generated by a restricted grammar.
  \end{lemma}
\begin{proof}
It is enough to prove the claim for an arbitrary language $L$ of the form (\ref{eq:linear}).  It is checked that $L$ is generated by the restricted grammar of index $k$ 
$G=\langle V, \Sigma, M, S \rangle$, 
  where $V = \{S, X_1, \ldots, X_k\}$ and $M$ is given by the matrices  
 \begin{itemize}

\item $(S\rightarrow X_1\cdots X_k)$,

\item % For every $X_i\in V_i$ and $1\leq i\leq k$, 
$(X_1 \rightarrow \varepsilon, \ldots, X_k \rightarrow \varepsilon)$,

\item $(S\rightarrow u_1^{{\bf v}_{01}}X_1\cdots u_k^{{\bf v}_{0k}}X_k)$, with ${{\bf v}_0=({\bf v}_{01}}, \ldots, {\bf v}_{0k})$,

\item For every $i=1, \ldots, \ell,$ $(X_1 \rightarrow u_1^{{\bf v}_{i1}}X_1,  \ldots, X_k \rightarrow u_k^{{\bf v}_{ik}}X_k)$ 
\\
with ${{\bf v}_i=({\bf v}_{i1}}, \ldots, {\bf v}_{ik}), \ 1\leq i\leq k$,
\end{itemize}
 where  the vectors  ${\bf v}_i, \ i=0, \ldots, k$, define (\ref{eq:linear-vectors}).  \qed 
  
 \end{proof}
  Proposition \ref{main-prop-gen-matr-gramm-bsl}   and Lemma \ref{gen-matr-gramm-bsl} 
 provide a new proof of the following. 
    \begin{corollary}{(\cite{dlt2014}, Prop. 21)}
\label{dlt2014-bis-bsl}
 If $L$ is  bounded semilinear then  
its closure $\Closure{\varphi(L)}$ is semialgebraic.
Furthermore, if the quantum automaton ${\cal Q}$ is rational, 
then the $(L, Q)$ Intersection problem is decidable.
   \end{corollary}
     \medskip

 \noindent
{\em Example \ref{first-example-matrix-gramm} (continued).} Let ${\cal Q}=(s,\varphi,P,\lambda)$ be a finite quantum automaton where
$\varphi :   \Sigma^* \longrightarrow O_n$ is the morphism (\ref{eq:main-morphism}) associated with ${\cal Q}$. % =(s,\varphi,P,\lambda)$. %   of the free monoid  $ \Sigma^*$  into the group $O_n$.
Recall that $\closure{({\cal A})}$ and  $\overline{{\cal A}}$ denote, respectively, the Euclidean and the Zariski closure of a set $\cal A$.
The set $\Closure{\varphi (L)}$ is then equal to  
\begin{equation}
\label{eq:first-example-matrix-gramm}
\Closure{\{\varphi (a)^n\varphi (b)^n\varphi (c)^n: n\in \N\}}
= \Closure{\Psi(\{\varphi (a)^n\oplus \varphi (b)^n\oplus \varphi (c)^n: n\in \N\})}
\end{equation}
%
%$$ \Closure{\{\varphi (a)^n\varphi (b)^n\varphi (c)^n: n\in \N\}}
%= \Closure{\Psi(\{\varphi (a)^n\oplus \varphi (b)^n\oplus \varphi (c)^n: n\in \N\})},$$
where $(\oplus)$ is the operation defined in (\ref{eq:sum-of-matrices}) and $\Psi$ is the map defined, for every $X, Y, Z\in O_n$, as
$\Psi (X\oplus Y\oplus Z):= XYZ$.
By Theorem \ref{th:fundamental}, the right-side  term of (\ref{eq:first-example-matrix-gramm}) rewrites as
$\Psi(\Closure{\{\varphi (a)^n\oplus \varphi (b)^n\oplus \varphi (c)^n: n\in \N\}})$, so that 
$$\Closure{\varphi (L)} = \Psi(\Closure{\{g\}^*}) = \Psi(\Closure{\langle\{g\}\rangle}),$$
where $g = \varphi (a)\oplus \varphi (b) \oplus \varphi (c)$.
By Theorem \ref{th:topological-closure-of-monoid}, $\Closure{\langle\{g\}\rangle} = \overline{{\langle\{g\}\rangle}}$ is algebraic. 
From this, by  Proposition~\ref{pr:product-of-blocks},
it follows that $\Closure{\varphi (L)} =\Psi( \overline{{\langle\{g\}\rangle}})$ is semialgebraic.

If the automaton $\cal Q$ is a rational, then, by applying Derksen's algorithm to $\{g\}$, one computes $\overline{{\langle\{g\}\rangle}}$ as an algebraic set,
which, in turn, implies that $\Psi( \overline{{\langle\{g\}\rangle}})$ is effective semialgebraic.
%More, if $\cal Q$ is a rational as a quantum automaton, then 
\medskip

   We observe that, not only,  Proposition \ref{main-prop-gen-matr-gramm-bsl}  provides a unitary scheme for the decidability results of \cite{dlt2014}  
   (i.e., Corollary~\ref{main:dlt2014} and  Corollary \ref{dlt2014-bis-bsl}) but it gives  a
    non trivial extension of them. Indeed
  the following example provides  languages, neither bounded nor context-free,  defined by restricted grammars, for which the 
  Intersection Problem is now decidable.
%   We finally observe that Proposition \ref{main-prop-gen-matr-gramm-bsl}  is a non trivial extension of Corollary~\ref{main:dlt2014} and  Corollary \ref{dlt2014-bis-bsl}. Indeed
%  the following example provides  languages, neither bounded nor context-free,  defined by restricted grammars.
   \begin{example}
    Let $L_k$, with $k\in \N$,  be the language over $\Sigma = \{a,b\}$ given by $L_k = \{u^k : u\in \Sigma^*\}$.
  It is easily checked that $L_k$ can be generated by the restricted grammar $G=\langle V, \Sigma, M, S \rangle$, 
  where $V = \{S, X_1, \ldots, X_k\}$, and $M$ is given by the matrix rules $(S\rightarrow X_1  \cdots X_k)$,  
 $(X_1 \rightarrow \varepsilon, \ldots, X_k\rightarrow~\varepsilon)$, and, for every $\sigma \in \Sigma$,
 $(X_1 \rightarrow \sigma X_1, \ldots, X_k \rightarrow \sigma X_k)$.
  Note that the very same argument used in Example \ref{first-example-matrix-gramm} allows one to verify that $\Closure{\varphi (L)}$
 is semialgebraic and effective semialgebraic in the case that the quantum automaton is rational. 
         \end{example}

 \subsection{The case of monoidal languages}
\label{subsec: the case of monoidal languages}
We start with the following definition that is instrumental for this section. We first recall that a
context-free linear grammar is said to be {\em minimal linear} if it has a unique variable.
 \begin{definition}
 \label{defo:monoidal grammar}
 A grammar $G=\langle V, \Sigma, P, S \rangle$ is said to be {\em monoidal of index $k$} if it is the composition
 $$G = {\cal G}_1 \circ {\cal G}_2 \circ \cdots \circ {\cal G}_k,$$
% of $k$ families of minimal linear grammars where, for each of them, every terminal production $\ (X\rightarrow u), \ u\in \Sigma^*,$\ is such that $u=\varepsilon$.
  of $k$ families of minimal linear grammars where every terminal production $\ (X\rightarrow u), \ u\in \Sigma^*,$\ is such that $u=\varepsilon$.
 \end{definition}
%
%
%
%
%%We start with the following definition that is instrumental for this section.
%%
%% \begin{definition}
%% \label{defo:monoidal grammar}
%% A grammar $G=\langle V, \Sigma, P, S \rangle$ is said to be {\em monoidal of index $k$} if it is the composition
%% $$G = {\cal G}_1 \circ {\cal G}_2 \circ \cdots \circ {\cal G}_k,$$
%% of $k$ families of linear grammars where the productions of $G$ satisfy the following conditions:
%%  \begin{itemize}
%% \item if $\ (X\rightarrow u)\in P, \, \ u\in \Sigma^* $ (terminal production)\,  then $u=\varepsilon$
%% 
%% 
%%\item if $\ (X\rightarrow uYv)\in P$, where $X, Y$ are distinct variables of a grammar in ${\cal G}_i, \ 1\leq i\leq k,$  then $u=v=\varepsilon.$
%%
%%%  \item if $\ (X\rightarrow uYv)\in P$, with $X \neq Y$, then $u=v=\varepsilon.$
%%
%%
%%% \item if $\ X\rightarrow uYv$ is a production of $G$ with $X \neq Y$, then $u=v=\varepsilon.$
%%%
%%%\item if $\ X\rightarrow u$ is a terminal production of $G$, then $u=\varepsilon$.
%%
%%\end{itemize}
%% \end{definition}
%
%We prove that, regardless of the length $k$ of the composition  (\ref{eq:intro-FI-gramm}) for $G$,
%if the Zariski closure over $\R$ of the monoids of cycles of the grammars of the lowest level  ${\cal G}_k$ of (\ref{eq:intro-FI-gramm}), are (algebraic) irreducible,
%then one can compute the Euclidean closure $\Closure{\varphi(L)}$ of $\varphi(L)$  (Proposition \ref{prop-main-monoidal}),
%
%
%
%
A language is said to be {\em monoidal} if it is generated by a monoidal grammar of index $k$, for some $k\in\N$.
The main result of this section is the following. 
     \begin{proposition}\label{prop-main-monoidal}   
  Let $(L,Q)$ be a pair where $Q$ is a rational quantum automaton
  and  $L$ is a language generated by a monoidal grammar $ G = {\cal G}_1 \circ {\cal G}_2 \circ \cdots \circ {\cal G}_k$.
      If the Zariski closures of the monoids of cycles
    of the grammars of ${\cal G}_k$ -- i.e., the lowest level  of $G$ --   are irreducible, then  $\Closure{\varphi(L)}$ is effective semialgebraic.
      \end{proposition}
      
      \begin{remark} Examples of monoids generated by a finite number of matrices whose Zariski closure is irreducible can be borrowed from \cite{gs21}. 
      In Theorem~1.1 of {\it loc.cit.},  the authors determine the Zariski closure of a cyclic matrix semigroup over $\mathbb C$. By the proof \cite[Corollary 17]{hopw2023bis}, the above result implies that if we consider the cyclic matrix semigroup generated by a real matrix with eigenvalues greater than $1$,
       its Zariski closure  is irreducible over $\mathbb R$.  Similarly, Example 3.8 from \cite{gs21} can be adapted to produce an example of 
    a   matrix semigroup $M$, generated by two matrices, such that the  
Zariski closure of $M$ is irreducible over $\mathbb R$.  %matrix semigroup generated by two matrices whose real Zariski closure is irreducible.
      \end{remark}
      
        We will do the proof of  Proposition \ref{prop-main-monoidal} for $k\leq 2$, i.e., for a monoidal grammar $G = {\cal G}_1 \circ {\cal G}_2$ of index $2$, 
        since, by (\ref{eq:bounded-lan-GinSpan}), the general case is treated with the same argument. 
    The proof  is technically %  involved and 
    structured in the following lemmata. 
%    
%    \noindent
   Also we find useful to recall a notation introduced in Section \ref{Sec:Irreducible sets}.
% \begin{definition}
% \label{defo:monoidalH}
%        {Given a subset $\cal A$ of the group  $O_{m}$ of orthogonal matrices (over $\R$), we write
%       $${\cal A} \in ({\cal H}),$$
%       if $\cal A$ is a finite union of irreducible, effective algebraic sets, every one of which
%       contains the identity matrix of $O_{m}$.}
%          \end{definition}
%      
%      
%      
%      
%      
%      
%      
%      
%      
%      
%      
       \medskip 
       
       \noindent
       {\bf Notation.} {\em Given a subset $\cal A$ of the group  $O_{m}$ of orthogonal matrices (over $\R$), we write
        \begin{equation}
 \label{defo:monoidalH}
{\cal A} \in ({\cal H})
\end{equation}
       if $\cal A$ is a finite union of irreducible, effective algebraic sets, every one of which
       contains the identity matrix $I$ of $O_{m}$.}
         \begin{lemma}\label{main-prop-main-monoidal-PART A} 
                     Let $G = {\cal G}_1$ be a monoidal grammar of index $1$ and let $L=L(G)$.
                              Under the assumption of  Proposition \ref{prop-main-monoidal},
there exist a set $H \in ({\cal H})$ and a regular map $\Psi$ such that  $\Closure{\varphi(L)}=\Psi(H).$
            Moreover $\Closure{\varphi(L)}$ is effective semialgebraic and contains $I$. 
%            Let $G = {\cal G}_1$ be a monoidal grammar of index $1$ and let $L=L(G)$. If the Zariski closure of all the monoids of cycles of $G$
%            are irreducible, effective algebraic sets, then there exist a set $H \in ({\cal H})$ and a regular map $\Psi$ such that  $$\Closure{\varphi(L)}=\Psi(H).$$
%            In particular $\Closure{\varphi(L)}$ is effective semialgebraic and contains $I$. 
%             \medskip
%       
%      \noindent
%      {\bf *** inoltre $\Psi(H)$ deve cont. I ***}\\
   \end{lemma}
    \begin{proof} 
    Set $G = \langle V, \Sigma, P, S \rangle$  and observe that $V=\{S\}$, so that $G$ has a sole monoid 
    of cycles $M_S$. Taking into account (\ref{eqfla:claim2-C}) and the form of the terminal productions of $G$,
    one has $\Closure{\varphi(L_S)}= \Psi(\Closure{M_S})$, where  $\Psi(X\oplus Y) := XY^{\scriptstyle T}.$ 
 From this and the fact that, by hypothesis,  $\Closure{M_S}$ is irreducible, the claim follows. % in this case. The basis of  induction is proved.
\qed   \end{proof}
%
%
%
% \begin{remark}      
% Note that the map $\Psi$ of the statement of Lemma \ref{main-prop-main-monoidal-PART A} does not depend upon the variable used as  axiom of $G$.
%% 
%% \noindent
%%{\bf  ***   gli $\Psi $ non dipendono dalle variabili di G}\\
%%  {\bf  ***  specificare meglio la FORMA di $\Psi$}
%%\\{\bf 
%%***  specificare meglio la FORMA di $\Psi ^{\prime\prime} $}
% \end{remark}
%
%
%
%
%
%
%    \medskip  \medskip  \medskip
%    
%     \noindent
%{\bf 
%***  introdurre considerazioni e notazioni su $G = {\cal G}_1 \circ {\cal G}_2$ }
%
%\medskip  \medskip  \medskip

%
%\noindent
Now assume $G = {\cal G}_1 \circ {\cal G}_2$ and let ${\cal G}_1=\langle V_1, \Sigma_1, P_1, S \rangle$ be the first grammar of the composition $G$. 
Set $V_1=\{S\}$.  
In the sequel, for every $A\in \Sigma_1$, we denote by $L_A$ the language generated by the grammar $G_A\in {\cal G}_2$. %  (cf Sec. \ref{Sec:Composition of grammars.}).
%Assume now that the first grammar of the composition $G = {\cal G}_1 \circ {\cal G}_2$ is ${\cal G}_1=\langle V_1, \Sigma_1, P_1, S \rangle$.
Let $\Sigma_1^* \times \Sigma_1^*$ be the multiplicative monoid of binary relations on $\Sigma_1^*$ and let $\Pi(O_n \oplus O_n)$ be the multiplicative monoid of subsets of $ O_n \oplus O_n$. % of $O_{2n} = O_n \oplus O_n$. 
Define now the morphism between the two monoids
  $$\zeta : \ \Sigma_1^* \times \Sigma_1^*\   \longrightarrow \ \Pi(O_n \oplus O_n)$$
%  $$\zeta : \ \Sigma_1^* \times \Sigma_1^*\   \longrightarrow \ \Pi(O_{2n})$$
as: for every $(u, v) \in  \Sigma_1^* \times \Sigma_1^*$, 
with $u= A_1 \cdots A_s, \ v= B_1\cdots B_t, \ A_i, B_i \in \Sigma _1$

  \begin{equation}
   \label{eq:main-prop-main-monoidal-PART B}
      \zeta (u, v) \ = \ ( \varphi (L_{A_1})  \cdots \varphi (L_{A_s}) \ \oplus \   \varphi (L_{B_1})^{\scriptstyle T} \cdots \varphi (L_{B_t})^{\scriptstyle T})
%   \zeta (u, v) \ = \ ( \varphi (L_{A_1})  \cdots \varphi (L_{A_s}), \ \   \varphi (L_{B_1})^{\scriptstyle T} \cdots \varphi (L_{B_t})^{\scriptstyle T})
   \end{equation}
where, for every $i=1, \ldots, t$, $\varphi (L_{B_i})^{\scriptstyle T}$ denotes the set of matrices
\begin{equation}\label{eq:set-trasp-matr}
 \{m^{\scriptstyle T} \ : \ m \in \varphi (L_{B_i}) \}.
\end{equation}

The following lemma holds.
  \begin{lemma}\label{main-prop-main-monoidal-PART B} 
    There exists an effective computable finite subset $\cal R$ of  $\ \Sigma_1 ^* \times \Sigma_1 ^*$ such that
 $M_S = \zeta ({\cal R}^*)$. 
%  For every $S\in V_1$, there exists an effective computable regular subset $\cal R$ of  $\ \Sigma_1 ^* \times \Sigma_1 ^*$ such that
% $M_S = \zeta ({\cal R})$. 
% There exists an effective computable regular subset $\cal R$ of  $\ \Sigma_1 ^* \times \Sigma_1 ^*$ such that, for every $S\in V_1$,
% $M_S = \zeta ({\cal R})$. % , where $\zeta$ is the map (\ref{eq:main-prop-main-monoidal-PART B}).
   \end{lemma}
   
     \begin{proof}
     Let us first define the  set   $$\widehat{C}_S = \{(\alpha, \beta)\in \Sigma_1^* \times \Sigma_1^* : (\alpha, \beta^\sim)\in C_S  \},$$
     where $\beta^\sim$ denotes the mirror image of $\beta$.
     Observe that $\widehat{C}_S$ is a submonoid of $\Sigma_1^* \times \Sigma_1^*$. Moreover one has $$M_S = \zeta (\widehat{C}_S).$$
     Indeed, let us check $M_S \subseteq \zeta (\widehat{C}_S).$  % By (\ref{eq:crucial-monoid}), 
If $\varphi(u)\oplus \varphi(v)^{\scriptstyle T}\in M_S$, then $S \displaystyle\mathop{\Rightarrow}^{*} u S v$. Since $G = {\cal G}_1 \circ {\cal G}_2$,  the last derivation can be rewritten as
 \begin{equation}
   \label{eq:main-prop-main-monoidal-PART B-sub 1}
   S \displaystyle\mathop{\Rightarrow}^{*} \alpha S \beta
      \end{equation}
      where (\ref{eq:main-prop-main-monoidal-PART B-sub 1}) is a derivation  of $ {\cal G}_1$ 
and
 \begin{equation}
   \label{eq:main-prop-main-monoidal-PART B-sub 2}
     \alpha \displaystyle\mathop{\Rightarrow}^{*} u, \quad \quad  \beta \displaystyle\mathop{\Rightarrow}^{*} v,
      \end{equation}
where  both   the derivations of (\ref{eq:main-prop-main-monoidal-PART B-sub 2}) are in the grammars of $ {\cal G}_2.$
Now set $\alpha= A_1A_2\cdots A_s,$ and $\beta = B_s B_{s-1}\cdots B_1$, with $A_i, B_i\in \Sigma_1 \cup\{\varepsilon\}, s\geq 0$.
From (\ref{eq:main-prop-main-monoidal-PART B-sub 2}), one then has
$$u = u_1\cdots u_s, \quad\quad v = v_s\cdots v_1,$$
where, for every $i=1,\ldots, s,$ $u_i \in L_{A_i},$ and $v_i \in L_{B_i}$. From this, one has
 \begin{equation}
   \label{eq:main-prop-main-monoidal-PART B-sub 3}
\varphi (u) \in \varphi (L_{A_1})\cdots \varphi (L_{A_s}), \quad\quad \varphi (v)^{\scriptstyle T} \in \varphi (L_{B_1})^{\scriptstyle T} \cdots \varphi (L_{B_t})^{\scriptstyle T}.
 \end{equation}
 Now observe that, by (\ref{eq:main-prop-main-monoidal-PART B-sub 1}), 
 $(\alpha, \beta)\in C_S$ so that $(\alpha, \beta^\sim)\in\widehat{C}_S$.   By (\ref{eq:main-prop-main-monoidal-PART B-sub 3}) the latter implies $\varphi(u)\oplus \varphi(v)^{\scriptstyle T}\in \zeta(\widehat{C}_S)$.
 Hence $M_S \subseteq \zeta (\widehat{C}_S)$.       The other inclusion is checked similarly.
 Finally observe that a finite set $\cal R$ of binary relations on $ \Sigma_1 ^*$ generating $\widehat{C}_S$ as a monoid,  
 can be effectively computed from the productions of ${\cal G}_1$.
   The lemma is proved. 
%  For every $S\in V_1$, there exists an effective computable regular subset $\cal R$ of  $\ \Sigma_1 ^* \times \Sigma_1 ^*$ such that
% $M_S = \zeta ({\cal R})$. 
      \qed   \end{proof}

%%The proof of the next lemma is easy and will be omitted.
%The proof of the next lemma is almost immediate.
%   \begin{lemma}\label{main-prop-main-monoidal-PART C-sub 1} 
%   Let $H_1, \ldots, H_m$ be subsets of an arbitrary monoid $\mbox{M}$, every one of which contains the identity $1$ of M. Then
%   $(H_1 \cup \ldots \cup  H_m)^* = (H_1\cdots H_m)^*$.
%   \end{lemma}
%   
%   \begin{proof} 
%   We prove the claim for $m=2$, i.e., $(A \cup B)^* = (AB)^*$ since the general case is similarly treated. 
%   The inclusion $(A \cup B)^* \supseteq (AB)^*$ is obvious. For the reverse inclusion, assume
%   $m=m_1\dots m_\ell, \ m_i\in A\cup B, \ 1\leq i \leq \ell$. Then $m$ can be written as
%   $m =a_1b_1\cdots a_\ell b_\ell$, where, for every $i=1,\ldots, \ell$
%   $a_i=m_i$ and $b_i=1$ if $m_i\in A$ while $a_i=1$ and $b_i=m_i$ if $m_i\in B$. 
%   \qed   \end{proof}
    
    \begin{lemma}\label{main-prop-main-monoidal-PART C} 
 $\Closure{M_S}\in ({\cal H}).$ In particular, $\Closure{M_S}$ is effective algebraic.
%     For every $A\in V_1$, $\Closure{M_A}\in ({\cal H}).$ In particular, $\Closure{M_A}$ is effective algebraic.
%      \medskip
%       
%      \noindent
%      {\bf *** rimettere a POSTO (i)\\
%%       *** rimettere a POSTO (iii)\\
% %       *** rimettere a Prop. 6\\
%       manca la osservazione, nei 3 casi, sulla unicita' di psi\\
%      {\bf *** inoltre $\Psi(H)$ deve cont. I ***}\\
%%      \bigskip
%%      
%%      \noindent
%%      *** BISOGNA aggiungere la dipendenza da $\Psi$		
%%      \\
%%      ***  la NON dipendenza di $\Psi$ dagli insiemi \\
%%      *** dominii delle funzioni regolari\\
%%            {\bf *** bisogna vedere se lo statement cosi come scritto e' CORRETTO -- 
%%      FORSE basta aggiungere una POSTILLA nel solo caso $\Closure{M_A}$? SI, basta una POSTILLA }
%}
        \end{lemma}
       \begin{proof}
        By  Lemma \ref{main-prop-main-monoidal-PART B}, 
         there exists an effective computable finite subset $\cal R$ of  $\ \Sigma_1 ^* \times \Sigma_1 ^*$ such that
 $M_S = \zeta ({\cal R}^*)$. 
   First let us  prove the following claim:
     \medskip
       
      \noindent
      {\bf Claim. }{\em Let $\cal R$ $=\{r\}$, with  $r\in \Sigma_1 ^* \times \Sigma_1 ^*$. Then $\Closure{\zeta ({\cal R})}$  
      contains $I$ and $\Closure{\zeta ({\cal R})} = \Psi(H)$ for some $H\in ({\cal H})$ (cf (\ref{defo:monoidalH}))
      and regular map $\Psi$.    }
         \medskip
       
      \noindent
%    \medskip
%       
%      \noindent
%      {\bf Claim. }{\em For every regular subset $\cal R$ of  $ \Sigma_1 ^* \times \Sigma_1 ^*$, $\Closure{\zeta ({\cal R})} = \Psi(H)$ for some $H\in ({\cal H})$ (cf (\ref{defo:monoidalH}))
%      and regular map $\Psi$ with $\Psi(H)$  containing $I$.  Further, if ${\cal R} = R^*$, then $\Closure{\zeta ({\cal R})} \in ({\cal H})$. }
%         \medskip
%       
%      \noindent
%We first show the claim for a finite set $\cal R$. We may suppose $\cal R$ $=\{r\}$, 
      Let us write $r=(\alpha,\beta)$ with $\alpha= A_1A_2\cdots A_s,\ $ $\beta = B_1 \cdots B_t$, $A_i, B_i\in \Sigma_1, \  s,t\geq 0$.
       We assume first that $\alpha$ and $\beta$ have the same length, i.e., $s=t$.
       By (\ref{eq:main-prop-main-monoidal-PART B}), the fact that $\zeta$ is a monoid morphism and Corollary \ref{cor:closure-of-product}, one gets
              $$\Closure{\zeta ({\cal R})} =  \ ( \Closure{\varphi (L_{A_1})}  \cdots \Closure{\varphi (L_{A_s})} \ \oplus \   \Closure{\varphi (L_{B_1})}^{\scriptstyle T} \cdots \Closure{\varphi (L_{B_s})}^{\scriptstyle T}),$$
%       $$\Closure{\zeta ({\cal R})} =  \ ( \Closure{\varphi (L_{A_1})}  \cdots \Closure{\varphi (L_{A_s})}, \ \   \Closure{\varphi (L_{B_1})}^{\scriptstyle T} \cdots \Closure{\varphi (L_{B_s})}^{\scriptstyle T}),$$
       taking into account that, for an arbitrary set $\cal A$ of orthogonal matrices,  for the set (\ref{eq:set-trasp-matr}), one has $\Closure{{\cal A}^{\scriptstyle T}} = \Closure{{\cal A}}^{\scriptstyle T}$. 
       By Lemma \ref{main-prop-main-monoidal-PART A}, the previous formula gives 
  $$\Closure{\zeta ({\cal R})} =  \ ( \Psi(H_1)  \cdots \Psi(H_s) \ \oplus \   \Psi(G_1)^{\scriptstyle T} \cdots \Psi(G_s)^{\scriptstyle T}),$$
%  $$\Closure{\zeta ({\cal R})} =  \ ( \Psi(H_1)  \cdots \Psi(H_s), \ \   \Psi(G_1)^{\scriptstyle T} \cdots \Psi(G_s)^{\scriptstyle T}),$$
  where $\Psi$ is a regular map and, for every $i=1, \ldots, s$, \ $H_i,  G_i \in ({\cal H})$.
   
   Since each $\Psi(H_i)$ and $\Psi(G_i)$ of the last formula, contain $I$ so holds for $\Closure{\zeta ({\cal R})}$.
  Set $H = H_1\oplus\cdots\oplus H_s \oplus G_1\oplus\cdots \oplus G_s$ and observe that, by Proposition \ref{pr:cartesian-product-irr-ET-other-OP}, $H\in ({\cal H})$. 
  Then $\Closure{\zeta ({\cal R})} =  \  \Psi^\prime (H)$, where $ \Psi^\prime$ is the regular map defined as
    $$ \Psi^\prime (X_1\oplus\cdots\oplus X_s \oplus Y_1\oplus\cdots \oplus Y_s) = (\Psi(X_1)\cdots\Psi(X_s)\oplus   \Psi(Y_1)^{\scriptstyle T} \cdots \Psi(Y_s)^{\scriptstyle T}).$$
%  $$ \Psi^\prime (X_1\oplus\cdots\oplus X_s \oplus Y_1\oplus\cdots \oplus Y_s) = (\Psi(X_1)\cdots\Psi(X_s), \   \Psi(Y_1)^{\scriptstyle T} \cdots \Psi(Y_s)^{\scriptstyle T}).$$
  This proves the claim  if $s=t$.
        Finally, if $s>t$, then we can write the word $\beta = B_1 \cdots B_t$ as $\beta = B_1 \cdots B_s$, with $B_j = \varepsilon$, for $j=t+1,\ldots, s$. Afterwards, 
 set $L_ {B_j} = \{\varepsilon\}$ and $\varphi(L_{B_j}) = \{I\}$, $t+1 \leq j\leq s$. Then one proceeds to prove  the claim, by repeating the argument above for $s=t$.
$\quad\quad\quad\quad\quad\quad\quad\quad\quad\bullet$ %$\quad\diamond$

\medskip
      \noindent   
     Since  $\zeta$ is a monoid morphism,  then one has 
$$\Closure{M_S}  = \Closure{\zeta ({\cal R}^*)}  = \Closure{\zeta ({\cal R})^*}.$$%  =  \Closure{( \Closure{\zeta ({R}}))^*} $$
%Observe that, from this, it trivially follows that $\Closure{\zeta ({\cal R})}$ contains $I$.

 \noindent 
Now taking into account that, by Corollary \ref{cor:closure-of-product}, for an arbitrary set $\cal X$, 
 \begin{equation}
 \Closure{{\cal X}^*} =  \Closure{ \Closure{{\cal X}}^*},
 \end{equation}
one has
$\Closure{M_S}   =  \Closure{ \Closure{ \zeta ({\cal R})}^*}. $
By Theorems \ref{th:semigroup-in-compact} and \ref{th:topological-closure-of-monoid}, one has 
  \begin{equation}\label{eq-claim-main-prop-main-monoidal-PART C-2}
  \Closure{ \Closure{ \zeta ({\cal R})}^*} = \overline{ \Closure{ \zeta ({\cal R})}^*}, % \Closure{ \Closure{ \zeta ({R})}^*}. $
%\Closure{ \Closure{ \zeta ({R})}^*} = \overline{\zeta ({R})^*}, % \Closure{ \Closure{ \zeta ({R})}^*}. $
\end{equation}
where $\overline{ \Closure{ \zeta ({\cal R})}^*}$ is the Zariski closure of $(\Closure{ \zeta ({\cal R})})^*$. 
Set ${\cal R} = \{r_1, \ldots, r_s\},$ with $r_i \in \Sigma_1^* \times \Sigma_1^*, s\geq 1$. 
Since, for every $i=1,\ldots, s$, $\Closure{ \zeta ({r_i})}$ contains the identity, by   Lemma \ref{main-prop-main-monoidal-PART C-sub 1}, one gets
 $\Closure{ \zeta ({\cal R})}^* = (\Closure{ \zeta ({r_1})}\cdots \Closure{ \zeta ({r_s})})^*$,
 which, thus, implies
 $$\Closure{M_S} = \overline{(\Closure{ \zeta ({r_1})}\cdots \Closure{ \zeta ({r_s})})^*}.$$
 
%
%%\overline{\zeta ({R})^*}$,
%%where  $ \overline{\zeta ({R})^*}$ is the Zarisky closure of $\zeta ({R})^*$. 
% 
%
%%(iii) By the fact that $\zeta$ is a monoid morphism and by Theorems \ref{th:semigroup-in-compact} and \ref{th:topological-closure-of-monoid}, one has 
%%$$\Closure{\zeta ({\cal R})}  = \Closure{\zeta ({R^*})}  = \Closure{\zeta ({R})^*}  = \overline{\zeta ({R})^*},$$
%%where  $ \overline{\zeta ({R})^*}$ is the Zarisky closure of $\zeta ({R})^*$. 
By the  claim, for every $i=1,\ldots, s$, 
$\Closure{ \zeta ({r_i})} = \Psi (H_i)$, where $H_i\in ({\cal H})$, and $\Psi$ is a regular map not depending upon $H_i$. 
Hence one gets
   \begin{equation}\label{eq-claim-main-prop-main-monoidal-PART C-2-fin}
\Closure{M_S} \ = \ \overline{( \Psi(H_1) \ \cdots \ \Psi (H_k))^*}.
 \end{equation}
%This fact and (\ref{eq-claim-main-prop-main-monoidal-PART C-2}) now  implies 
%$$\Closure{\zeta ({\cal R})}  =  \overline{\Psi(H)^*}. $$
%%where $\Psi$ is a regular map and $H\in ({\cal H})$. 
%Since $H\in ({\cal H})$, we can write
%$H = \bigcup _{i =1 }^k\ H_i$
%%$$H = \bigcup _{i \in \cal I}\ H_i$$ 
%as  a finite union of irreducible effective algebraic sets $H_i$, every one of which containing the identity matrix~$I$. 
%This, together with Lemma \ref{main-prop-main-monoidal-PART C-sub 1},  implies 
%%\begin{equation}\label{eq-claim-main-prop-main-monoidal-PART C-2}
%%$$ \overline{\zeta ({R})^*} \ = \ \overline{( \Psi(H_1) \ \cdots \ \Psi (H_k))^*}.$$
%%$$\Closure{\zeta ({\cal R})}  \ = \ \overline{( \Psi(H_1) \ \cdots \ \Psi (H_k))^*}.$$
%$$\Closure{\zeta ({\cal R})}  \ = \ \overline{( \Psi(H_1) \ \cdots \ \Psi (H_k))^*}.$$
%\end{equation}
Observe  that in  (\ref{eq-claim-main-prop-main-monoidal-PART C-2-fin}) we can write $\Psi (H_1) \ \cdots \ \Psi (H_k)$ as % \\
$\widehat{\Psi}(H_1\oplus \cdots \oplus H_k)$,
 where $\widehat{\Psi}$ is the regular map defined as 
 $\widehat{\Psi}(X_1\oplus \cdots \oplus X_k) = \Psi(X_1) \cdots \Psi (X_k)$.
 Hence we get
%  \Closure{ \Closure{ \zeta ({\cal R})}^*} = \overline{ \Closure{ \zeta ({\cal R})}^*}, % \Closure{ \Closure{ \zeta ({R})}^*}. $
%\Closure{ \Closure{ \zeta ({R})}^*} = \overline{\zeta ({R})^*}, % \Closure{ \Closure{ \zeta ({R})}^*}. $
$$ %\Closure{\zeta ({\cal R})}  
\Closure{M_S}  = \overline{( \widehat{\Psi}(H))^*},$$
  where, again, $H= (H_1\oplus \cdots \oplus H_k) \in ({\cal H})$ as one can easily check.
Finally $\Closure{M_S}\in ({\cal H})$  follows by applying Proposition \ref{pr:Psi-Ir}.
  % This completes the proof. 
   \qed   \end{proof}

             We are now able to prove the main result of this section.
       \medskip
       
      \noindent
      {\bf Proof of Proposition \ref{prop-main-monoidal}:} 
            As remarked before,  it is enough to prove the claim for $k\leq 2$ since, by (\ref{eq:bounded-lan-GinSpan}), the general case is treated with the same argument. 
       Set $G = \langle V, \Sigma, P, S \rangle$ and observe that, for every $A\in V$, the fact that $\Closure{M_A}$ is effective algebraic comes from 
       Lemmata \ref{main-lm-lin:dlt2014}  and \ref{main-prop-main-monoidal-PART C}.
     The claim now follows from Proposition \ref{main-prop-dlt2014}. 
%
%
%
%      As remarked before, we prove the claim for $k\leq 2$ since, by (\ref{eq:bounded-lan-GinSpan}), the general case is treated with the same argument. 
%       Let $G = \langle V, \Sigma, P, S \rangle$.
%     By Proposition \ref{main-prop-dlt2014}, it is enough to show that,  for every $A\in V$,  
%      $\Closure{M_A}$ is effective algebraic. 
%      This condition comes from the hypothesis on $G$ if 
%       $k=1$ and from Lemma \ref{main-prop-main-monoidal-PART C} if $k=2$.
%  
%      
%      
%
%      Let $G=\langle V, X, P, S \rangle$ be the grammar with $L=L(G)$. We now prove that, for each $S\in V$, $\Closure{\varphi(L_S)}$ is effective semialgebraic.
%      For this purpose, recall that, by hypothesis,
%      $G = {\cal G}_1 \circ {\cal G}_2$. Let ${\cal V}_1$
%      be the set of variables of  ${\cal G}_1$ and ${\cal V}_2$ be the set of variables of the grammars in ${\cal G}_2$, respectively. 
%      Since ${\cal V}_1 \cap {\cal V}_2=\emptyset$, either $S\in {\cal V}_1$ or $S\in {\cal V}_2$. In the latter, the claim immediately comes from Lemma~\ref{main-prop-main-monoidal-PART A}.
      \qed
      
      \medskip
      
      \noindent
      By Propositions \ref{pr:union-prod-phi} and  \ref{pr:meta-resuositionlt},  immediate corollaries of Proposition \ref{prop-main-monoidal} are the following.

    \begin{corollary}{}
\label{cor-main-monoidal-1}
 If the pair $(L, Q)$ satisfies the hypothesis of Proposition \ref{prop-main-monoidal}, then   the $(L, Q)$ Intersection problem is decidable.
   \end{corollary}

    \begin{corollary}{}
\label{cor-main-monoidal-2}
Let $(L,Q)$ be a pair where $Q$ is a rational quantum automaton
  and  $L$ is a finite union $L=\bigcup _{i\in {\cal I}} L_i,$ of monoidal languages,
 every one of which is such that the corresponding pair $(L_i, Q)$ satisfies the hypothesis of Proposition \ref{prop-main-monoidal}.
 Then the  $(L, Q)$ Intersection problem is decidable.
    \end{corollary}

   \section{Concluding remarks and open problems}\label{sec: final comments}
    We finally address some questions arising  from this study. 
    \medskip

\noindent 
%{\bf [$\alpha$]\;}    We design a procedure to decide the {\sc $(L, Q)$ Intersection} problem for languages generated by monoidal grammars
% and irreducible automata. Such procedure involves the computation of the Zariski closure of a group $G$ of matrices that is
%{\em algebraically presented}, i.e. $G$ admits a set of generators that is algebraic itself. 
%It may be of interest to investigate conditions assuring the extension of the algorithms of  \cite{derksen,npssw2022} for such groups.     
{\bf [$\alpha$]\;}    
One could extend  Proposition \ref{prop-main-monoidal} to a broader class of quantum automata. 
In this context, the  construction of the decidability procedure would  require the computation of the Zariski closure of a group $G$ of matrices that is
{\em algebraically presented}, i.e. $G$ admits an infinite set of generators that is algebraic itself. 
It may be of interest to investigate conditions ensuring the extension of the algorithms of  \cite{derksen,hopw2023,npssw2022} for such groups. 
%We design a procedure to decide the { $(L, Q)$ Intersection} problem for languages generated by monoidal grammars
% and irreducible automata. One could extend this  result   for a broader class of quantum automata. 
%In this context, the  construction of the decidability procedure involves the computation of the Zariski closure of a group $G$ of matrices that is
%{\em algebraically presented}, i.e. $G$ admits a set of generators that is algebraic itself. 
%It may be of interest to investigate conditions assuring the extension of the algorithms of  \cite{derksen,hopw2023,npssw2022} for such groups. 
\medskip

\noindent 
{\bf [$\beta$]\;} It is open whether the $(L, Q)$ Intersection problem is decidable for context-free languages. We conjecture that the  answer to this problem is   negative. 
 An approach may be to reduce the  problem to a so called {\em ``birdie problem"} \cite{ullian} (cf also \cite{Hu}, Ch. 8). In such a problem, 
 assuming the feasibility of the computation of the Zariski closure of the groups above, one gets, as a consequence, 
 the decidability of some unsolvable problem. %  on such languages.
 
\medskip

\noindent 
{\bf [$\gamma$]\;} The {$(L, Q)$ Intersection} problem is decidable for bounded semi-linear languages. It may be of interest to get extensions
 of this  result for the broader class of languages accepted by the  reversal bounded non deterministic counter machines \cite{ibarra-main}. This extension could exploit 
a characterization of such languages provided by Ibarra in  \cite{ibarra} in terms of some special combinatorial systems called RLGMC-grammars
(Right-Linear Grammar with Multi-Counters). % Essentially, a 
A RLGMC-grammar is a regular grammar that associates with each
  derivation a weight, i.e. a vector in the additive monoid $\N^d$. Then a derivation is considered successful if the corresponding weight belongs to the monoid generated by 
  $\bf 1^d$ $=(1,1,\ldots,1)$. 
    In this case,  the construction of the formula  that makes the problem decidable, is reduced to the computation of
   the Zariski closure of the intersection of two finitely generated monoids of matrices, say $M_1$ and $M_2$. 
   One can compute $\overline{M_1}$,  $\overline{M_2}$ and, consequently,  $ \overline{M_1}\cap \overline{M_2}$. However, in general, the inclusion
       $\overline{M_1\cap M_2} \subseteq \overline{M_1}\cap \overline{M_2}$ is strict and the intrinsic difficulty to 
 {\em control} the accumulation points of the set -- i.e., to exclude those  not coming from the effective functioning of the machines --  makes  such a computation a non trivial task.

\vskip5pt
 
\noindent{\bf\large Acknowledgments}
\vskip5pt
The authors would like to thank Mima Stanojkovski for
useful discussion on her paper \cite{gs21}.
%
%{The authors would like to  thank Mima Stanojkovski  for  stimulating discussion on some constructive aspects of  finitely
%generated monoids  whose Zariski closure is irreducible.
%%We thank  the reviewers for their suggestions which helped us to improve the presentation of the content and of the results of this work.

% \newpage

   \begin{center}
\bf \Large Appendix 
\end{center}
We recall the construction  of the context-free grammar that generates
 the image $\theta (L)$ of a context-free language $L$ over the alphabet $\Sigma_1$ 
 under a {\em context-free substitution} $\theta : \Sigma_1^* \rightarrow \Pi(\Sigma_2^*)$,
 i.e.  a {$\cal F$-substitution}, with $\cal F$=CFL.
 
   We follow {\em verbatim} the argument of \cite{Ber79}, Sec. II.2. 
    \noindent
   Let ${\cal G}_1=\langle V_1, \Sigma_1, P_1, S \rangle$ be the context-free  grammar that generates $L$ % with $\Sigma_1=X$,   
   and   let 
   $${\cal G}_2 = \{G_x \ : \ x\in \Sigma_1\}$$
   be the family of context-free grammars where, for every $x\in \Sigma_1$,   
  $$G_x=\langle V_x, \Sigma_2, P_x, S_x \rangle$$ is the grammar that generates $\theta(x)$.
    One may assume that, for all $x\in \Sigma_1$, 
  $V_1\cap V_x=\emptyset$ and, for every $x, y \in \Sigma_1,\ x \neq y$, $V_x\cap V_y=\emptyset$.   
   Let be the set $${\cal V}_2 = \{S_x : x\in \Sigma_1\}$$ and let the  copy morphism 
        \begin{equation}\label{eq:cm}
           \mbox{c} : (V_1\cup \Sigma_1)^* \rightarrow (V_1\cup {\cal V}_2)^*, 
  %   \mbox{c} : (V_1\cup \Sigma_1)^* \rightarrow (V_1\cup \{S_x : x\in \Sigma_1\})^*, 
           \end{equation}
   defined as: $\mbox{c} (z)=z$, if $z\in V_1$, and $\mbox{c} (z)=S_z$, if $z\in \Sigma_1$.
   With respect to $\mbox{c}$, consider the grammar $G_{\mbox{c}}$ that generates $\mbox{c} (L)$. 
   Such a grammar is defined as $$G_{\mbox{c}} =\langle V_1, \{S_x : x\in \Sigma_1\}, P_{\mbox{c}}, S \rangle,$$ 
   where $P_{\mbox{c}} = \{A \rightarrow \mbox{c}(\alpha) : (A \rightarrow \alpha)\in P _1\}$.
     Finally, define   
     \begin{equation}\label{eq:bounded-lan-GinSpan-Ex}
     H=\langle W, \Sigma_2, Q, S \rangle 
      \end{equation}
 as the context-free grammar where 
   $$W = V_1 \ \cup \ \bigcup _{x\in \Sigma_1}\ V_x, \quad \quad Q = P_{\mbox{c}} \ \cup \ \bigcup _{x\in \Sigma_1}\ P_x.$$
 Such grammar $H$ is called the {\em composition} of  ${\cal G}_1$ and  ${\cal G}_2$ and denoted by
  \begin{equation}\label{eq:bounded-lan-GinSpan-comp-rk2}
     G = {\cal G}_1 \circ {\cal G}_2.
       \end{equation}
       
%          \begin{remark}\label{rmk2}
%          By a minor abuse of notation, up to the copy morphism (\ref{eq:cm}), one may identify $\Sigma_1$ with ${\cal V}_2$. In particular ${V}_1\cap {\cal V}_2 = \emptyset.$
%        \end{remark}

 The following lemma holds.
 \begin{lemma}%\label{}
 If $G$ is the grammar (\ref{eq:bounded-lan-GinSpan-comp-rk2}), then $\theta (L)=L({G})$.
  \end{lemma}
  A {\em composition of $k$ families of  context-free grammars} $({\cal G}_1,   \ldots,  {\cal G}_k),$ with $k\geq 1$ % and $\card({\cal G}_1)=1$, 
will be  the context-free  grammar
 $$G = {\cal G}_1 \circ {\cal G}_2 \circ \cdots \circ {\cal G}_k,$$
 obtained by iterating $(k-1)$ times the operation (\ref{eq:bounded-lan-GinSpan-comp-rk2}) of composition of grammars  on the families
% of context-free grammars  
${\cal G}_1,   \ldots,  {\cal G}_k$. %  (if the operation is well defined).

 \end{document}